\documentclass[fleqn,usenatbib]{mnras}

\usepackage{newtxtext,newtxmath}
\usepackage{physics}

\usepackage[T1]{fontenc}

\DeclareRobustCommand{\VAN}[3]{#2}
\let\VANthebibliography\thebibliography
\def\thebibliography{\DeclareRobustCommand{\VAN}[3]{##3}\VANthebibliography}
\def\beq{\begin{equation}}
\def\eeq{\end{equation}}

\def\softwarenamestyle[#1]{\textsc{#1}}

\defcitealias{moTwophaseModelGalaxy2023}{Paper-I}

\usepackage{graphicx}	
\usepackage{amsmath}	
\usepackage[dvipsnames]{xcolor}
\usepackage{enumitem}
\usepackage{url}
\usepackage{braket}
\usepackage{threeparttable}
\usepackage{hyperref}
\usepackage{multirow, tabularx}

\def\Mpc{\,{\rm Mpc}}
\def\mpc{\, h^{-1}{\rm {Mpc}}}

\def\kpc{\, h^{-1}{\rm {kpc}}}

\def\Msun{{\rm M_\odot}}
\def\msun{\, h^{-1}{\rm M_\odot}}

\title[Size-mass relation of galaxies]{A two-phase model of galaxy formation: II. The size-mass relation of dynamically hot galaxies}

\author[Chen, Mo and Wang]
{Yangyao Chen,$^{1,2}$\thanks{E-mail: yangyaochen.astro@foxmail.com}
Houjun Mo,$^{3,4}$
and Huiyuan Wang$^{1,2}$ 
\\
$^{1}$School of Astronomy and Space Science, University of Science and Technology of China, Hefei, Anhui 230026, China\\
$^{2}$Key Laboratory for Research in Galaxies and Cosmology, Department of Astronomy, University of Science and Technology of China, Hefei, Anhui 230026, China \\
$^{3}$Department of Astronomy, University of Massachusetts, Amherst MA01003, USA\\
$^{4}$Tsung-Dao Lee Institute, Shanghai Jiao Tong University, Shanghai 200240, China\\
}

\date{Accepted XXX. Received YYY; in original form ZZZ}

\pubyear{2023}

\begin{document}
\label{firstpage}
\pagerange{\pageref{firstpage}--\pageref{lastpage}}
\maketitle

\begin{abstract}
In Paper-I we developed a two-phase model to connect 
dynamically hot galaxies (such as ellipticals and bulges) 
with the formation of self-gravitating gas clouds (SGCs) associated with 
the fast assembly of dark matter halos. Here we explore the 
implications of the model for the size-stellar mass relation of dynamically hot galaxies. 
Star-forming sub-clouds resulting from the fragmentation of the turbulent SGC
inherit its spatial structure and dynamical hotness, 
producing a `homologous' relation, $r_{\rm f}\approx\, 100 r_{\rm bulge}$, 
between the size of a dynamically hot galaxy ($r_{\rm bulge}$)
and that of its host halo assembled in the fast regime ($r_{\rm f}$), 
independent of redshift and halo mass. 
This relation is preserved by the `dry' expansion driven by dynamical 
heating when a galaxy becomes gas-poor due to inefficient cooling, 
and is frozen due to the stop of bulge growth during the slow assembly 
regime of the halo. The size-stellar mass relation is thus 
a simple combination of the galaxy-halo homology 
and the non-linear stellar mass-halo mass relation.
Using a set of halo assembly histories we reproduce all properties 
in the observed size-mass relation of dynamically hot galaxies, 
including the flattening in the low-mass end and the upturn in the 
massive end. 
The prediction matches observational data currently available to 
$z \approx 4$, and can be tested in the future at higher $z$. 
Our results indicate that the sizes of dynamically hot galaxies are produced 
by the dissipation and collapse of gas in halos to establish 
SGCs in which stars form.
\end{abstract}




\section{Introduction}
\label{sec:introduction}

In the current $\Lambda$CDM cosmology, galaxies form and evolve in the gravitational 
potential wells of cold dark matter (CDM) halos. As gravitational systems in quasi dynamical 
equilibrium, galaxies are expected to obey the virial theorem at any given time, 
which relates the gravitational potential energy and the kinetic energy of a dynamical system. 
For a galaxy, the details of the relation depend on the mass distribution and 
velocity structure. From the perspective of galaxy formation and evolution, 
these are determined by the assembly of different mass components into the galaxy, 
energy dissipation of the star-forming gas, and the eventual support 
of the structure (e.g. by random motion or rotation) in the dynamical equilibrium. 
Clearly, the size-mass relation, which relates two of the key 
quantities in the virial theorem,  carries important information about 
the formation and evolution of galaxies.

For a dynamically cold galaxy, i.e. that dominated by a stable disk, 
analytical investigations suggested that its size is determined by the size and spin of 
its host dark matter halo \citep[e.g.][]{moFormationGalacticDiscs1998}. 
This prediction is roughly in agreement with observational data 
\citep[e.g.][]{moFormationGalacticDiscs1998,
shenSizeDistributionGalaxies2003MassSizeRelation,
somervilleRelationshipGalaxyDark2018}.
On the other hand, the formation theory of dynamically hot galaxies, especially for 
elliptical galaxies, remains uncertain. Theoretical investigations
based either on empirical \citep{
shenSizeDistributionGalaxies2003MassSizeRelation,
bezansonRelationCompactQuiescent2009,
newmanCanMinorMerging2012,
huangRelationsSizesGalaxies2017,
somervilleRelationshipGalaxyDark2018}, 
semi-analytical \citep{ciottiImportanceDryWet2007} or numerical 
\citep{oserTwoPhasesGalaxy2010,
hopkinsDiscriminatingPhysicalProcesses2010,
hopkinsMergersBulgeFormation2010,
oserCosmologicalSizeVelocity2012,
mccluskeyDiskSettlingDynamical2023} 
approaches suggested a variety of formation scenarios for dynamically hot 
galaxies, such as in-situ star formation in dissipative and fast collapse
of the gas component \citep[e.g.][]{somervilleRelationshipGalaxyDark2018}, 
gas-rich (wet) and gas-poor (dry) mergers 
\citep[e.g.][]{shenSizeDistributionGalaxies2003MassSizeRelation,
ciottiImportanceDryWet2007,
bezansonRelationCompactQuiescent2009,
oserTwoPhasesGalaxy2010,
hopkinsDiscriminatingPhysicalProcesses2010,
hopkinsMergersBulgeFormation2010,
oserCosmologicalSizeVelocity2012,
newmanCanMinorMerging2012}, 
expansions due to 
significant mass loss driven by supernova (SN) and active galactic nucleus (AGN)
feedback \citep[e.g.][]{fanDramaticSizeEvolution2008,
bezansonRelationCompactQuiescent2009,
hopkinsDiscriminatingPhysicalProcesses2010}, 
and secular evolution 
\citep[e.g.][]{kormendySecularEvolutionFormation2004,
sahaEffectBarsTransient2010,
sahak.DynamicalEvolutionBulge2012,
mccluskeyDiskSettlingDynamical2023}.  
However, when and how these processes shape the size-mass evolution of dynamically 
hot galaxies are still poorly understood. There are several reasons for this.
First, observational data of high-$z$ galaxies required to 
constrain models directly are difficult to obtain. 
Second, the dissipative nature and violent collapse of the gas component 
in the high-$z$, gas-rich Universe, coupled with the expected feedback processes, 
are difficult to follow numerically. Third, mergers of galaxies 
are diverse in gas richness, mass distribution, orbital properties,
and the exchange of energy and angular momentum between stars and dark matter, 
making their effects difficult to quantify. Finally, feedback processes are
extremely difficult to model reliably, even with the state-of-the-art 
hydro simulations. 
The difficulty partly arises from the diverse sources of feedback whose 
generation, propagation, and coupling with the gas are not fully understood. 
For example, the EAGLE \citep{schayeEAGLEProjectSimulating2015} and 
NIHAO simulations \citet{blankNIHAOXXIIIntroducing2019} both implement a purely thermal 
subgrid model for AGN feedback. In contrast, the TNG simulation introduces an 
additional kinetic-mode channel that seems crucial 
in quenching massive galaxies \citep{weinbergerSimulatingGalaxyFormation2017,
weinbergerSupermassiveBlackHoles2018}. The SIMBA simulation further incorporates 
an X-ray channel for AGN feedback \citep{daveSimbaCosmologicalSimulations2019}
that appears to be important in producing the bimodal distribution in 
the stellar-to-halo mass relation of galaxies \citep{cuiOriginGalaxyColour2021}.
Another source of the difficulty is the limitation of computational power 
in resolving the large dynamic range of physical scales involved in galaxy 
formation. For example, zoom-in simulations such as FIRE
\citep{hopkinsFIRE2SimulationsPhysics2018,hopkinsFIRE3UpdatedStellar2023}
and SMUGGLE \citep{liEffectsSubgridModels2020} can incorporate detailed 
radiative feedback from stars, which is suggested to be critical in regulating 
small-scale gas properties. Numerical and physical convergences of stellar 
feedback are thus achieved only when a certain resolution threshold is reached
\citep{dengSimulatingIonizationFeedback2024}.

Observationally, the size-mass relation of dynamically hot galaxies 
presents rich features, indicating that many processes may be relevant. 
\citet{shenSizeDistributionGalaxies2003MassSizeRelation} analyzed the 
size-mass relation using the SDSS data at $z \approx 0.1$, 
and found that the relation can be well-fitted by a power law with an index of 
$0.56$ for elliptical galaxies with stellar mass 
$M_* \gtrsim 10^{10} \Msun$. A plausible formation scenario for these 
ellipticals was then suggested, in which the mass and size growths 
are both driven by repeated dry mergers with moderately bound orbits. This 
scenario thus indicates an energy transfer from merging galaxies to dark matter.
Using a one-parameter form for the energy transfer calibrated to match 
the observed size-mass relation, this scenario can reproduce the observed 
power-law index. However, later observations \citep{vandokkumConfirmationRemarkableCompactness2008,
bezansonRelationCompactQuiescent2009,vanderwel3DHSTCANDELSEvolution2014,
vandokkumFormingCompactMassive2015,newmanCanMinorMerging2012,
millerNewViewSize2019,mowlaMassdependentSlopeGalaxy2019} indicated that 
the size-mass relation evolves significantly with redshift:  
at given mass, the size decreases by $\approx 0.4\,$dex from $z=0$ to 
$z \approx 2.5$. If small ellipticals are progenitors of massive ones, 
the small ellipticals must obey a size-mass relation 
with lower amplitude, and the underlying size-mass relation must be 
steeper than the size-mass relation observed at a given redshift.
This will lead to a strong evolution in the slope of the size-mass relation, 
unless the binding energy is finely tuned to depend on the redshift of the 
observed descendant. Meanwhile, the repeated merger scenario does not explain the 
flattening of the size-mass relation for lower-mass ellipticals 
($M_* < 10^{10}\Msun$) seen in \citet{shenSizeDistributionGalaxies2003MassSizeRelation} 
and confirmed subsequently in other observations 
\citep[e.g.][]{grahamEllipticalDiskGalaxy2013,langeGalaxyMassAssembly2015,
mowlaMassdependentSlopeGalaxy2019,nedkovaExtendingEvolutionStellar2021}.
In addition, for the most massive elliptical galaxies, observations also 
found evidence for a non-linear upturn in the size-mass relation 
\citep{bezansonRelationCompactQuiescent2009,langeGalaxyMassAssembly2015, grahamEllipticalDiskGalaxy2013,
nedkovaExtendingEvolutionStellar2021}, which needs to be explained by theory as well. 

With a proper transformation of the observed size-mass relation 
by relating the stellar mass with the halo mass using abundance-matching, 
\citet{huangRelationsSizesGalaxies2017} found that the relation between halo 
virial size and galaxy size is linear for early-type galaxies, and the slope is quite 
independent of redshift all the way to $z \sim 3$
\citep[see also][]{somervilleRelationshipGalaxyDark2018}.
This galaxy-halo `homology' is inspiring, as it suggests that the 
size-mass relation of dynamically hot galaxies may originate from 
some simple processes that relate galaxies and dark matter halos.

In our first paper of this series \citep[][hereafter Paper-I]{moTwophaseModelGalaxy2023}
we provided a framework to model the formation of dynamically hot galaxies and 
their supermassive black holes (SMBHs). The key to this framework is the two-phase 
assembly of dark matter halos. The early, fast phase of assembly is associated with 
rapid cooling and collapse of gas to form a 
self-gravitating gas cloud (referred to as SGC).
High-density gas sub-clouds formed through the fragmentation of the turbulent SGC
move roughly ballistically without being significantly affected by the cloud-cloud 
collision and drag force from the surroundings, producing a dynamically 
hot system of protogalaxy. Stars that form in sub-clouds inherit the 
hot dynamics of the gas, producing a dynamically hot stellar system. 
In the late phase of slow assembly, the gas supply is reduced
because of the slow accretion of new gas, ineffective cooling 
and feedback from SN and AGN. As demonstrated in \citetalias{moTwophaseModelGalaxy2023}, without much fine-tuning, 
the model can reproduce a number of observations, including the correlations among 
SMBH mass, stellar mass of galaxies, and halo mass, as well as their evolution over 
the cosmic time. Relevant to this paper, the model predicts that the 
size of a dynamically hot galaxy is determined by the size of the SGC
that forms the galaxy. 

In this paper, we explore the size-mass relation of dynamically
hot galaxies predicted in the framework set up in \citetalias{moTwophaseModelGalaxy2023}. 
We will demonstrate that the physical conditions of galaxy formation 
expected from current cosmology naturally predicts a halo-galaxy homology.
We will show that our model can not only recover all the main features 
seen in the observed size-mass relation at different redshift, but also 
can provide a clear picture regarding their physical origins.  
In what follows, we begin with a brief description of our model
(\S\ref{sec:model}). We then present our model
predictions for the size-mass relation of dynamically hot galaxies
and compare them with observations (\S\ref{sec:interp-obs}). 
Finally, we summarize our main findings, compare our model with a 
number of other models in the literature, and discuss the potential of 
testing our model by future observations and simulations (\S\ref{sec:summary}).

\section{The model}
\label{sec:model}

\subsection{Coevolution of dark matter halos, galaxies and SMBHs}
\label{ssec:coevolution}

The model we developed in \citetalias{moTwophaseModelGalaxy2023} is built on dark matter halo assembly histories, 
$M_{\rm v}(z)$, defined to be the virial mass at redshift $z$ in the main branch 
of a halo. Here, the virial mass is defined to be $M_{\rm 200c}$, the total 
mass enclosed in a radius, $r_{\rm v}\equiv r_{\rm 200c}$, within which the mean density 
is 200 times the critical density of the Universe at the redshift of interest.
As discussed by, e.g. \citet{wechslerConcentrationsDarkHalos2002} 
and \citet{zhaoGrowthStructureDark2003}, the assembly of a halo 
can be divided into two phases: an early fast accretion phase where the 
halo concentration is roughly a constant, $c \approx c_{\rm f} = 4$,
and a late slow accretion phase where matter piles up in the outer part of 
the halo so that the concentration increases with time.
Thus, for each assembly history, we can apply a specially tailored root-finding 
algorithm to obtain the transition redshift, $z_{\rm f}$, when the specific growth 
rate of the halo drops below a certain fraction of the Hubble expansion rate
at that redshift. Specifically, we use the solution of    
$\dot{V}_{\rm v}/V_{\rm v} = \gamma_{\rm f}H(z)$ to determine $z_{\rm f}$,   
where $V_{\rm v}$ is the virial velocity of the halo, $H(z)$ is the Hubble parameter, 
and $\gamma_{\rm f}$ is a constant parameter. We refer to the history at 
$z \geqslant z_{\rm f}$ ($z < z_{\rm f}$) as the fast (slow) assembly phase
of the halo. To avoid confusion, we use a subscript `f' to denote
virial properties of the halo assembled in the fast phase, and use 
`v' to denote virial properties evaluated at the redshift of interest. 
For example, in the fast assembly phase, we have 
$M_{\rm f} = M_{\rm v}$ and $r_{\rm f}=r_{\rm v}$, and in the slow assembly phase, we 
have $M_{\rm f} < M_{\rm v}$ and $r_{\rm f} < r_{\rm v}$.

In the fast assembly phase, the gaseous halo collapses and cools 
rapidly, forming a system of SGC that is turbulent and subsequently
fragments into gas sub-clouds.
These sub-clouds are expected to be dense so that they 
move ballistically without being significantly affected by cloud-cloud 
collision and drag force from the surroundings. Stars formed in sub-clouds
inherit their dynamics, producing a dynamically hot stellar system
reminiscent of a bulge or an elliptical galaxy, and sub-clouds with low 
specific angular momentum can sink to the halo center to feed the SMBH. 
SN and AGN feedback heats and/or ejects gas from the SGC, 
which may suppress the star formation and SMBH growth. 
As discussed in \S3.1 of \citetalias{moTwophaseModelGalaxy2023} based on the 
cooling diagram,  the cooling time scale, $t_{\rm cool}$, of the shock-heated halo gas 
significantly exceeds the free-fall time scale, $t_{\rm ff}$, in massive halos with 
$M_{\rm v} \gtrsim 10^{13} \msun$, implying an inefficient gas supply to the galaxy.
This, combined with the AGN feedback, regulates star formation and the growth of 
the SMBH, and eventually leads to the quenching of both.

Based on the above consideration, we developed a set of equations
to describe the radiative cooling, star formation, SMBH growth, feedback processes, 
and gas evolution in individual halos to follow the evolution of different 
mass components, such as the stellar mass ($M_*$), the mass of the SMBH ($M_{\rm bh}$), 
and the gas mass ($M_{\rm g}$). 
In particular, within a given time interval in the 
fast assembly phase, the increases of the stellar mass and SMBH mass are 
assumed to be related to the increase of the halo mass by
\begin{equation} \label{eq:delta-m-star}
    \Delta M_*  = \epsilon_{\rm *,f} F_{\rm agn} F_{\rm sn} F_{\rm cool} f_{\rm B} \Delta M_{\rm f}
\end{equation}
and
\begin{equation} \label{eq:delta-m-bh}
\Delta M_{\rm bh}   = \alpha_{\rm cap} { M_{\rm bh} \over M_{\rm g} } 
            F_{\rm en} F_{\rm agn} F_{\rm sn} F_{\rm cool} f_{\rm B} \Delta M_{\rm f}  \,,
\end{equation}
where $f_{\rm B}$ is the universal baryon fraction; $F_{\rm cool}$ determines the 
fraction of the accreted gas that can cool down to feed the turbulent SGC; 
$F_{\rm sn}F_{\rm agn}$ is a term to describe the fraction of cooled gas that 
is not affected by the SN and AGN feedback in its ability to form sub-clouds in the SGC;
$\epsilon_{\rm *,f}$ is the star formation efficiency in a sub-cloud;
$F_{\rm en}$ is an enhancement factor describing the effects of `positive' 
feedback in driving turbulence and producing low-angular-momentum sub-clouds 
to be captured by the SMBH; and $\alpha_{\rm cap}M_{\rm bh}/M_{\rm g}$ is the 
fraction of sub-clouds that can be captured by the SMBH, derived by 
using the broad distribution of specific angular momentum in SGC.

One critical step in our model is to determine the detailed functional forms
of the operators in equations (\ref{eq:delta-m-star}) and (\ref{eq:delta-m-bh}).
Based on our understanding of these processes, we adopt the following forms, 
\begin{align}\label{eq:def-f-sn}
    F_{\rm cool} (M_{\rm f}\vert M_{\rm cool},\beta_{\rm cool}) & =
    {1 \over 1+(M_{\rm f}/M_{\rm cool})^{\beta_{\rm cool}}}     \,;\\
    F_{\rm sn}(V_{\rm g}\vert \alpha_{\rm sn}, \beta_{\rm sn}, V_{\rm w})  & = 
    {\alpha_{\rm sn}  +(V_{\rm g}/V_{\rm w})^{\beta_{\rm sn}}
    \over 
    1 +(V_{\rm g}/V_{\rm w})^{\beta_{\rm sn}}}           \,;\\    
    F_{\rm agn} (M_{\rm bh}, M_{\rm g}, V_{\rm g}\vert \alpha_{\rm agn}) 
     & =1- {\alpha_{\rm agn} M_{\rm bh} c^2 \over M_{\rm g} V_{\rm g}^2}   \,; \\     
    F_{\rm en} (M_{\rm f}\vert \alpha_{\rm en}, \beta_{\rm en}, M_{\rm en})
    & = {\alpha_{\rm en} + (M_{\rm f}/M_{\rm en})^{\beta_{\rm en}}
    \over 1+ (M_{\rm f}/M_{\rm en})^{\beta_{\rm en}}}\,.
\end{align}
For $F_{\rm cool}$, the threshold $M_{\rm cool} \approx 10^{13} \msun$ determines 
the regime where radiative cooling becomes ineffective, 
and $\beta_{\rm cool} > 0$ characterizes the sharpness of the transition 
so that a larger value of $\beta_{\rm cool}$ gives a more rapid decrease of 
the cooling with halo mass.
The supernova feedback term is assumed to depend on the maximum circular velocity 
of the halo, $V_{\rm g} \equiv V_{\rm max}$, and $\alpha_{\rm sn}$ and $V_{\rm w}$ 
characterize the strength of the dependence so that the fraction of star-forming gas is reduced 
to $\alpha_{\rm sn}$ when $V_{\rm g} \ll V_{\rm w}$.
The quantity $\alpha_{\rm agn}$ describes the AGN feedback 
energy injected and coupled to the SGC, and the effect is controlled by the 
binding energy of the SGC. 
The enhancement of SMBH growth by `positive' stellar feedback is expected to be 
significant only in low-mass halos with $M_{\rm f} \lesssim M_{\rm en}$, 
where stellar feedback effectively drives the turbulent motion of gas and 
gas cooling efficiently replenishes the cold gas, leading to turbulent 
motion of gas clouds within the galaxy \citep[see, e.g.][]{maSelfconsistentProtoglobularCluster2020,
hopkinsWhatCausesFormation2023,shiHyperEddingtonBlackHole2023}. 
The parameter $\alpha_{\rm en} > 1$ represents the enhancement factor applied 
to the SMBH growth rate for $M_{\rm f} \ll M_{\rm en}$. The parameter 
$M_{\rm en}$ characterizes the mass scale at which positive feedback becomes unimportant, 
while the parameter $\beta_{\rm en}$ controls the transition rate.
All these parameters have physically presumable 
prior ranges, but their precise values have to be calibrated by observations. 
As described in \citetalias{moTwophaseModelGalaxy2023}, we only use the $M_*$-$M_{\rm v}$ relation for central 
galaxies \citep{yangEVOLUTIONGALAXYDARK2012}
and the $M_{\rm bh}$-$M_{\rm *, bulge}$ relation for early-type galaxies
\citep{grahamAppreciatingMergersUnderstanding2023}, obtained at $z \approx 0$ to calibrate 
our model without further fine-tuning.  

In this paper, we use a halo sample and the main-branch assembly histories generated  
by Monte Carlo methods based on the mass functions of \citet{murrayHMFHaloMass2014} and 
the assembly history sampling method of \citet{hearinDifferentiableModelAssembly2021}, 
respectively. At each desired redshift, we generate a sample of 1024 halos 
with mass uniformly distributed in logarithmic space in the range of 
$10^{9.3} \msun \leqslant M_{\rm v} \leqslant 10^{15.5} \msun$, and   
we randomly sample the main branch assembly history for each halo.
Halo profile is assumed to have the NFW form \citep{navarroUniversalDensityProfile1997} with a 
concentration $c = c_{\rm f} \approx 4$ in the fast accretion phase \citep[e.g.][]{zhaoGrowthStructureDark2003}.  
The characteristic radius of the profile is denoted as $r_{\rm s}$, and the inner radius, 
$r_{\rm f} \approx c_{\rm f} r_{\rm s}$, encloses the mass, $M_{\rm f}$, that 
has been assembled in the fast phase. The $\Lambda$CDM cosmology 
adopted has $h=0.6774$, $(\Omega_{\rm M,0},\,\Omega_{\rm B,0},\,\Omega_{\rm \Lambda, 0})
=(0.3089,\,0.0486,\,0.6911)$, and a spectral index $n=0.9667$ and amplitude given by 
$\sigma_8=0.8159$. These are consistent with the Planck2015 results \citep{planckcollaborationPlanck2015Results2016}.
All the radii presented in this paper are physical values, rather than 
comoving values.

For detailed modeling of all the mass components, the list of model parameters, 
and the halo sampling and weighting method, please refer to the Low-$z_{\rm f}$ variant described in 
\citetalias{moTwophaseModelGalaxy2023}. In all predictions, a log-normal scatter with $\sigma = 0.2$ dex is added 
to $r_{\rm bulge}$ \citep{vanderwel3DHSTCANDELSEvolution2014}, 
$M_{\rm *, bulge}$ \citep{conroyModelingPanchromaticSpectral2013},
and $M_{\rm bh}$ \citep[as a rough estimate of the minimal scatter; e.g.][]{zhuangEvolutionaryPathsActive2023},
to mimic observational uncertainties. 
We note that our conclusion is robust against the choice of the adopted scatter.

\subsection{The buildup of galaxy-halo homology in the wet stage}
\label{ssec:size-wet}

A direct prediction of our bulge formation scenario based on the formation 
of SGC is the galaxy-halo homology, namely a constant galaxy size-halo size ratio 
during the gas-rich (wet) stage of fast accretion when the cooling of halo gas is effective.
If we denote the fraction of gas mass in SGC as $f_{\rm gas} \equiv M_{\rm gas}/M_{\rm f}$, 
a contraction of halo gas by a factor of $f_{\rm gas}$ will result in 
self-gravitating and subsequent fragmentation of the SGC. Thus, the size of an SGC,
and the characteristic size of the stellar bulge formed in it, 
is proportional to $f_{\rm gas}\times r_{\rm f}$. In general, we can write:
\begin{equation} \label{eq:def-r-bulge}
    r_{\rm bulge} 
    = f_{\rm r} f_{\rm gas} r_{\rm f}
    = f_{\rm r} (1 - f_{\rm ej}) f_{\rm B} r_{\rm f} \,.
\end{equation}
Here, $f_{\rm ej}$ is the fraction of the cooled gas that is ejected from 
the SGC by feedback. In \citetalias{moTwophaseModelGalaxy2023}, we found that,  as long as cooling is effective, 
self-regulations by stellar and SMBH growth produce a roughly constant 
fraction, among the total available gas, that is affected by the feedback 
(see their Fig.~6). The rapid dissipation due to the high gas fraction 
and the rapid change of the gravitational potential in the fast assembly phase 
is expected to adjust the energy distribution of the gas affected by the 
feedback according to the gravitational potential, regardless of the 
details of the feedback process. 
Consequently, $f_{\rm ej}$, the fraction of feedback-affected gas that can 
escape from the SGC, is also a constant defined by the tail of the energy distribution. 
The value of $f_{\rm ej}$ can be determined by considering the boundary 
condition at the end of the fast assembly phase. At this moment, the gas 
within the galaxy is about to transit to a stable disk once the strong disturbance 
from rapid halo assembly stops. The stability condition, as suggested by,
e.g. \citet[][see their \S3.2]{moFormationGalacticDiscs1998}, implies 
that $f_{\rm gas} \approx \lambda_{\rm gas} \approx 0.04$, 
where $\lambda_{\rm gas}$ represents the effective spin of 
the gas. The value of 0.04 is similar to the median halo 
spin of the universal spin 
distribution \citep[e.g.][]{bullockUniversalAngularMomentum2001,
maccioConcentrationSpinShape2007,bettSpinShapeDark2007}, 
and is also similar to that obtained for observed galaxy disks 
\citep[e.g.][]{
shenSizeDistributionGalaxies2003MassSizeRelation,
somervilleExplanationObservedWeak2008,
desmondTullyFisherMasssizeRelations2015,
burkertANGULARMOMENTUMDISTRIBUTION2016,
somervilleRelationshipGalaxyDark2018}.
Given that the universal 
baryon fraction $f_{\rm B} = 0.16$, approximately $3/4$ of the gas is expected to 
be ejected to reduce $f_{\rm gas}$ to the stable value. Thus, we set 
$f_{\rm ej} = 3/4$ as the fiducial value in our model.
Finally, $f_{\rm r}$ is a constant factor specific to the definition 
of the bulge size. As found previously 
\citep[e.g.][]{millerNewViewSize2019,mowlaMassdependentSlopeGalaxy2019,
suessHalfmassRadii70002019,vanderwelStellarHalfmassRadii2024}, 
adopting different definitions of radius can result in 
different size-mass relations. The factor $f_{\rm r} \lesssim 1$, with 
its value calibrated by observational data, is included to take account of the 
arbitrariness in the definition.  We use the observational results obtained by 
\citet{shenSizeDistributionGalaxies2003MassSizeRelation}  
for present-day elliptical galaxies with stellar mass 
$M_* \approx 5\times 10^{10} \msun$ to calibrate $f_{\rm r}$,  
and we obtain $f_{\rm r} = 0.26$. Their choice of radius is the $r$-band half-light 
radius, which is commonly adopted in the literature.
Taking all the factors as described above, equation~\eqref{eq:def-r-bulge} 
approximately gives 
\begin{equation}
    r_{\rm bulge} \approx 0.01\, r_{\rm f}\,,
\end{equation}
or
\begin{equation}
    r_{\rm f} \approx 100\, r_{\rm bulge}\,,
\end{equation}
regardless of the halo mass and redshift, as long as the halo is in 
the fast assembly phase and radiative cooling is effective.
We note that the definition of $r_{\rm bulge}$ is not unique in observations 
and may have systematic differences depending on factors such as sample selection, 
the method used to fit the light profiles and to handle the wavelength dependence. 
As shown by \citet{vanderwelStellarHalfmassRadii2024}, for quiescent galaxies, 
the 2-D half-stellar-mass radius is approximately 0.1–0.14 dex smaller than 
the optical half-light radius for galaxies with $M_* > 10^{10} \Msun$ at $z < 2.5$ (see their \S2.4), 
while the 3-D (deprojected) half-stellar-mass radius is about 0.05 dex larger 
than the 2-D half-stellar-mass radius (see their \S2.6). 
Consequently, the systematic difference between the 3-D half-stellar-mass 
radius and the 2-D optical half-light radius is largely canceled out,
leaving these two radii consistent with each other within 0.1 dex. 
Hence, $r_{\rm bulge}$ modeled and calibrated here can 
serve as a proxy for both the 2-D half-light radius in the optical band and 
the 3-D half-stellar-mass radius.

\begin{figure*} \centering
    \includegraphics[width=0.9\textwidth]{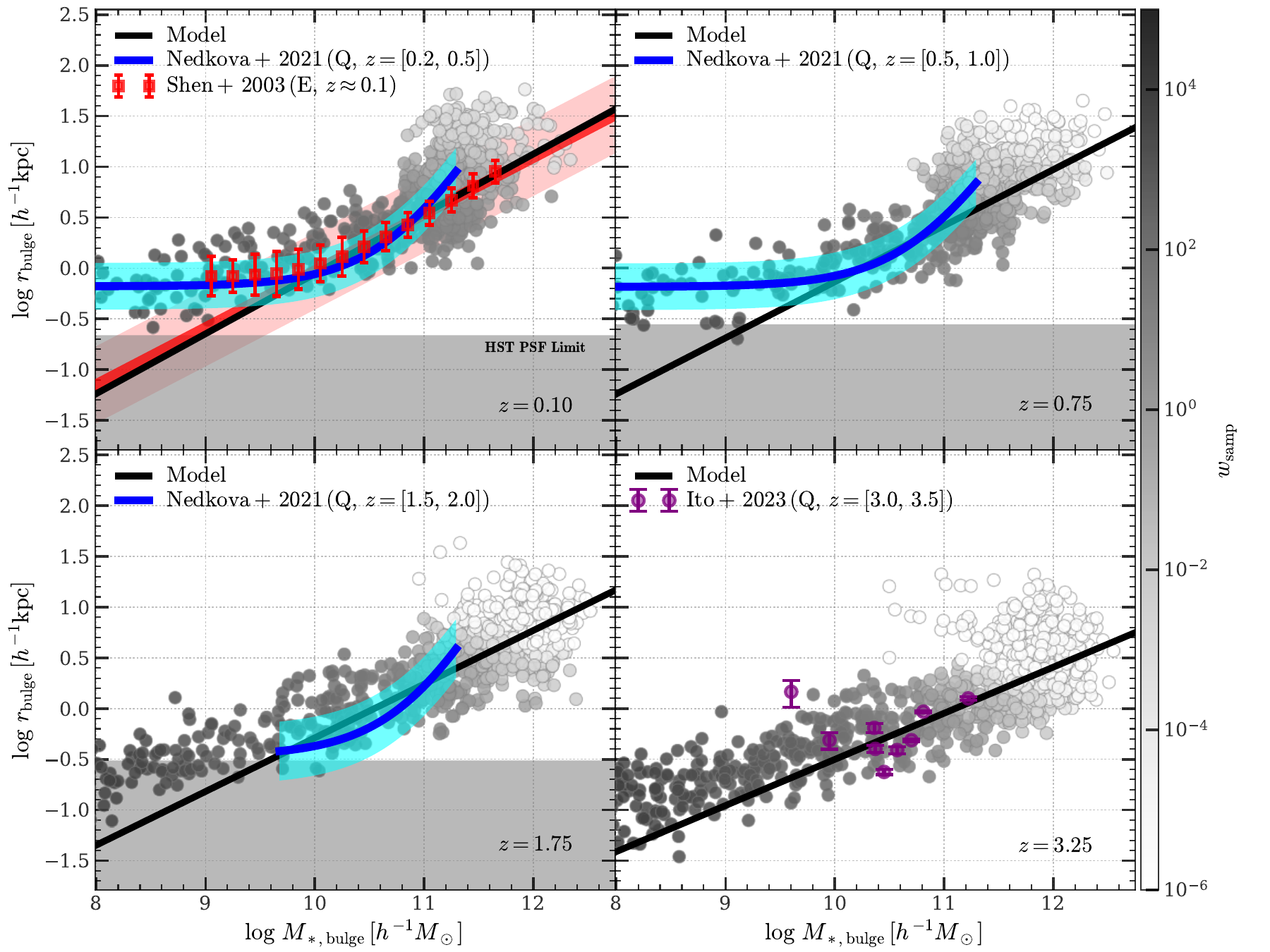}
    \caption{
        Size-mass relation of dynamically hot galaxies.
        Results at $z=0.1$, $0.75$, $1.75$ and $3.25$ are shown 
        in {\bf four panels}, respectively. Galaxies with bulge mass fraction 
        $f_{\rm bulge} > 0.75$ are included.
        In each panel, {\bf gray scatter points} are our model prediction 
        for the bulge size $r_{\rm bulge}$ and bulge stellar mass 
        $M_{\rm *, bulge}$ of individual galaxies, color-coded by the sampling 
        weight ($w_{\rm samp}$) according to the colorbar.
        {\bf Black solid line} is a linear fit for galaxies with
        $M_{\rm *, bulge} > 10^{10} \msun$.
        In the first panel. 
        {\bf Red squares} with error bars are size-mass relation 
        obtained by \citet[][see their Fig.~8]{shenSizeDistributionGalaxies2003MassSizeRelation}
        for SDSS early-type (E) galaxies defined by $^{0.1}(g-r)>0.7$ at $z\approx 0.1$,
        where $r$-band Sersic half-light radius is used as galaxy size.
        Here, a constant mass-to-light ratio is used to convert their $r$-band Sersic
        absolute magnitude to stellar mass.
        {\bf Red line} is their linear fit for high-mass galaxies (see their Table 1), 
        which has a logarithmic slope of $0.56$. 
        In each of the first three panels,
        {\bf blue curve} is the size-mass relation obtained
        by \citet[][see their Fig.10 and Table 2]{nedkovaExtendingEvolutionStellar2021}
        using HST images for quenched (Q) galaxies defined by 
        a UVJ color-color cut at the redshift range indicated in the legend.
        The 5000~Å half-light radius is used as their galaxy sizes. Gray shaded 
        area in the bottom of the panel indicates the size below 
        ${\rm FWHM}_{\rm F160W}/2$ of the point spread function.
        The red and blue shading areas indicate their residuals of fitting.
        In the last panel, {\bf purple dots} with errorbars show the measurements 
        obtained by \citet{itoSizeStellarMass2023} 
        using images from a number of recent JWST surveys
        for quenched galaxies selected by multiple color-based criteria 
        at $3 \leqslant z \leqslant 3.5$. Here 
        the 5000~Å effective radius along the semi-major axis is used as their 
        galaxy size. Only the `good-fit' galaxies are included (see their Table 2).
        This figure indicates that our model can reproduce the observed
        size-mass relations, including the slope of high-mass galaxies, 
        the non-linearity, and the evolution over the cosmic time 
        (see \S\ref{ssec:compare-with-obs} for the details).
    }
    \label{fig:scatters}
\end{figure*}

\begin{figure} \centering
    \includegraphics[width=0.975\columnwidth]{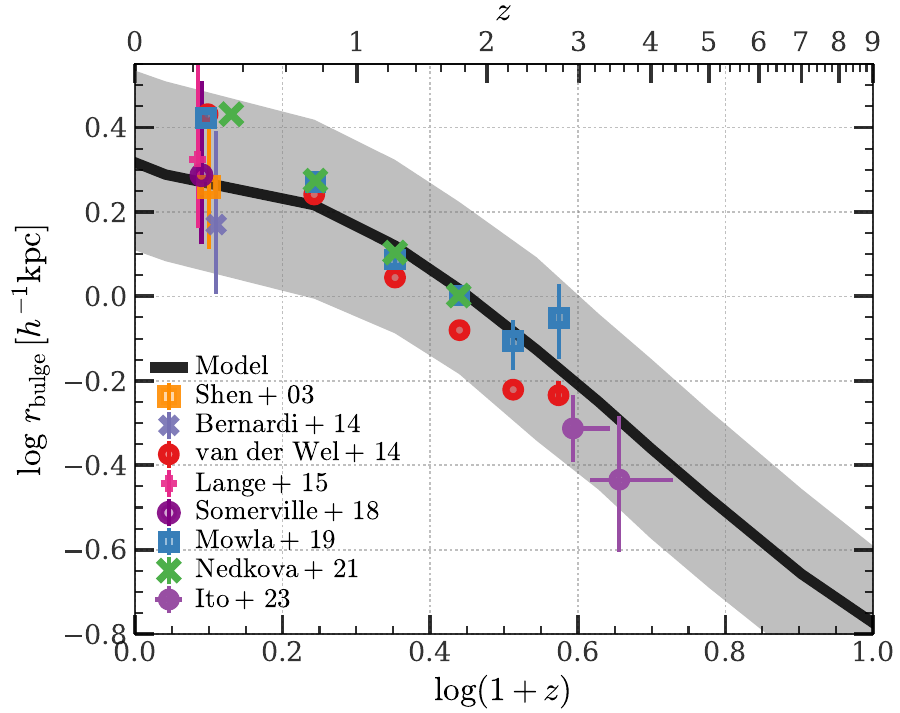}
    \caption{
        Size evolution of dynamically hot galaxies at 
        $M_{\rm *, bulge} = 5\times 10^{10} \Msun$ up to redshift $z=9$.
        {\bf Black line} shows the median size at each given redshift
        predicted by our model, with the gray shading area indicating 
        the 1-$\sigma$ (16\%-84\%) quantiles.
        {\bf Colored markers} are results at $z \lesssim 4 $ obtained 
        by different observations
        \citep{shenSizeDistributionGalaxies2003MassSizeRelation,
        bernardiSystematicEffectsSizeluminosity2012,
        vanderwel3DHSTCANDELSEvolution2014,
        langeGalaxyMassAssembly2015,
        somervilleRelationshipGalaxyDark2018,
        mowlaMassdependentSlopeGalaxy2019,
        nedkovaExtendingEvolutionStellar2021,
        itoSizeStellarMass2023}. See \S\ref{ssec:compare-with-obs} 
        for a detailed description of this figure.
    }
    \label{fig:evolution}
\end{figure}

\begin{figure*} \centering
    \includegraphics[width=\textwidth]{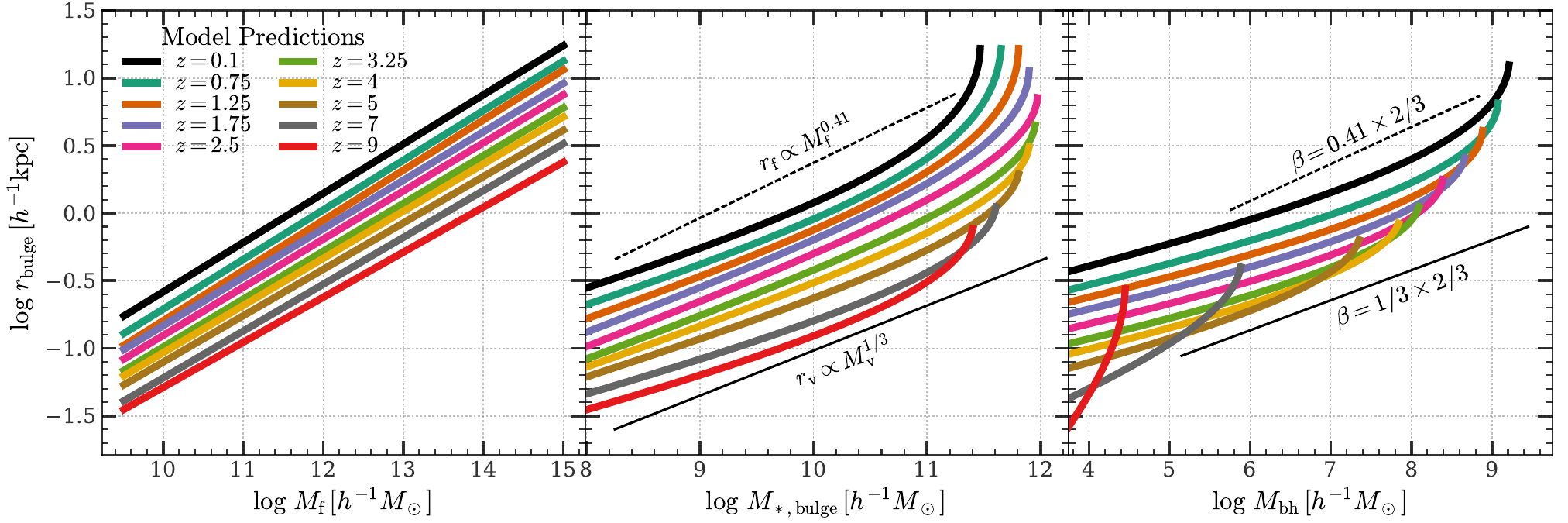}
    \caption{
        Size-mass relation of dynamically hot galaxies predicted by our model 
        up to $z=9$. 
        The {\bf left panel} shows the bulge size ($r_{\rm bulge}$)-halo mass ($M_{\rm f}$)
        relations.
        Curves with {\bf different colors} show the fitted relation at 
        different redshifts, as labeled in the legend.
        The {\bf center panel} shows the bulge size ($r_{\rm bulge}$)-
        bulge mass ($M_{\rm *,bulge}$) relations (referred to as size-mass relations). 
        The {\bf black dashed line} indicates the $r_{\rm f}$-$M_{\rm f}$ relation 
        (arbitrarily scaled) of halos fitted from simulated halo at $z=0$,
        with a power-law index $\beta_{\rm f,max}=0.41$.
        The {\bf black solid line} indicates the $r_{\rm v}$-$M_{\rm v}$ relation
        (arbitrarily scaled) at arbitrary redshift, which by definition has a 
        power-law index of $\beta_{\rm v}=1/3$.
        The {\bf right panel} is similar but shows the bulge size-SMBH
        mass ($M_{\rm bh}$) relations. Black dashed and solid lines are obtained 
        by scaling the power-law indices $\beta_{\rm v}$ and $\beta_{\rm f,max}$ 
        with $2/3$ to account for the $M_{\rm bh}$-$M_{\rm *,bulge}$ scaling relation
        for high-mass systems ($M_{\rm *,bulge}\gtrsim 10^{10} \msun$).
        See \S\ref{ssec:compare-with-obs} for a detailed description of this 
        figure and \S\ref{ssec:interp-size-mass} for a physical interpretation.
        For the list of fitted parameters, see Table~\ref{tab:parameters}.
    }
    \label{fig:fitting}
\end{figure*}

\subsection{The preservation of the galaxy-halo homology in the dry stage}
\label{ssec:size-dry}

For massive halos that exceed the cooling threshold ($M_{\rm cool}$) in their 
fast assembly phase, the gas supply is cut off and the galaxy becomes a gas-poor 
(dry) system that enters into the regime of quenching. 
However, since halo is still in the fast assembly regime, the gravitational 
potential of the halo is still changing rapidly.

One process that may be responsible for the late growth of giant ellipticals is 
gas-poor minor merger 
\citep[][]{naabMinorMergersSize2009,bezansonRelationCompactQuiescent2009,
hopkinsDiscriminatingPhysicalProcesses2010}. 
As small galaxies are in general less bound by self-gravity than massive ones, 
minor mergers inject positive binding energy into the merger remnant, 
which makes the remnant expand without changing the total stellar mass significantly.
However, even with relatively strong energy injection, minor mergers 
alone may not be able to explain the rapid size growth of giant ellipticals 
\citep{hopkinsMergersBulgeFormation2010}. Recent hydrodynamic simulations 
show that there may be more channels for the energy injection, especially during the 
fast phase of halo assembly. These channels, collectively referred to as `dynamical heating', 
include all processes that can repeatedly scatter stars.  The sources of 
scattering can be very diverse. Giant molecular clouds, spiral arms, global 
asymmetries, feedback-driven outflows, cosmic accretion, mergers, and close 
interaction with satellites can all have their impacts
\citep[see, e.g.][for a summary]{mccluskeyDiskSettlingDynamical2023}.

Despite the details, the effects of dynamical heating in a fast accreting halo 
may be simple to describe. The halo profile in the fast assembly phase is 
found to be almost universal, an NFW profile with a nearly constant 
concentration parameter $c\approx c_{\rm f}$ 
\citep{zhaoGrowthStructureDark2003,luOriginColdDark2006}. Only the size 
and amplitude of the halo density profile change with time. This is a homology 
in the halo population. Thus, the dynamical heating due to frequent and repeated 
scattering in the fast regime has a global effect that causes the expansion of 
the halo as a whole without changing its self-similar structure (homology). 
Consequently, all the collisionless components, including the central galaxy and
the inner region ($r<r_{\rm s}$) of the halo, experience similar evolution.  
The bulge size-halo size ratio, therefore, remains unchanged during the 
evolution in this regime.  Such a growth thus preserves the galaxy-halo homology 
that has built up in the wet stage.

\bigskip
Putting the above modeling together, we obtain a prediction for the bulge size-halo size 
relation for dynamically hot galaxies, which is  
$r_{\rm f} \approx 100\, r_{\rm bulge}$ throughout the fast assembly phase. 
The physical origin of this relation is clear and simple:
\begin{enumerate}[topsep=0pt]
    \item the homology is built up from {\em wet contraction} to form  
    self-gravitating gas clouds that are turbulent and fragment rapidly to form  
    dense star-forming sub-clouds, and
    \item for massive halos that can enter the gas-poor stage, the homology is 
    preserved by {\em dry expansion} driven by dynamical heating.
\end{enumerate}

\section{Interpretation of the observed size-mass relation for dynamically hot galaxies}
\label{sec:interp-obs} 

\subsection{Model predictions and comparisons with observational data}
\label{ssec:compare-with-obs}

The gray dots in Fig.~\ref{fig:scatters} show our model prediction 
for the size and mass of individual dynamically hot central galaxies. 
Here, we define dynamically hot galaxies as those with the bulge mass fraction
$f_{\rm bulge} = M_{\rm *, bulge}/M_* > 0.75$, and we have tested that the results are robust
against the choice of the threshold. Results at four different redshifts, from 
$z=0.1$ to $z=3.25$, are shown in the four panels, respectively.
Each dot is color-coded according to its sampling weight, $w_{\rm samp}$,
defined to be the number of occurrences of such a galaxy 
in a cubic box with a side length $l_{\rm box} = (75\mpc)^3 \approx (100 \Mpc)^3$.
The black line is a linear (in log-log space) fitting for all galaxies with 
$M_{\rm *, bulge} \geq 10^{10} \msun$, and represents a power-law 
with an index $\approx 0.6$ and an amplitude that decreases systematically 
with increasing redshift. For comparison, red squares with error bars in the first panel 
are the size-mass relation obtained by \citet{shenSizeDistributionGalaxies2003MassSizeRelation}
for local elliptical galaxies using the SDSS data. The red line shows a linear 
fit to high-mass galaxies ($M_* \gtrsim 10^{10} \msun$), with the shaded area 
indicating the residual of the fitting. The power-law index of this relation is 
$0.56$. Their result for high-mass galaxies has nearly the same amplitude as our 
results, as is expected from the fact that it is used to  
calibrate the only free parameter $f_{\rm r}$ in our model. However, our prediction for the 
slope is also consistent with the data, indicating that our model can 
match the observed mass dependence of galaxy size. In the same figure, we also 
show the results for quenched galaxies obtained by \citet{nedkovaExtendingEvolutionStellar2021}
from HST images, and by \citet{itoSizeStellarMass2023} from JWST images.
At $M_{\rm *, bulge} \approx 10^{10}$--$10^{11} \msun$, the model predictions match the 
observational results well, indicating that our model can
reproduce the observed redshift evolution of the size-mass relation.

An interesting feature of the model prediction is the non-linearity 
at both the low- and high-mass ends. At $M_{\rm *, bulge} \leqslant 10^{10}\msun$, 
the predicted relation is flattened, with an asymptotic power index smaller than 
$1/3$ (see below). This flattening was first seen in 
\citet{shenSizeDistributionGalaxies2003MassSizeRelation}
at $ z \approx 0.1 $, as shown by the red symbols in the first panel of 
our figure, and confirmed by subsequent observations at different redshifts
\citep[e.g.][]{grahamEllipticalDiskGalaxy2013,langeGalaxyMassAssembly2015,
mowlaMassdependentSlopeGalaxy2019,
millerNewViewSize2019,nedkovaExtendingEvolutionStellar2021}. 
At the high-mass end, $M_{\rm *, bulge} > 10^{11} \msun$,
the predicted relation shows an upturn, which is dominated by massive and 
rare objects. This upturn is seen in all bands analyzed by 
\citet{langeGalaxyMassAssembly2015},  and seen at all redshifts
below $z=2$ in \citet{nedkovaExtendingEvolutionStellar2021}.
To deal with the non-linearity, these authors adopted a double-power-law model  
to fit the observed size-mass relation over the full stellar mass 
range covered by the data. The results
of \citet{nedkovaExtendingEvolutionStellar2021} are shown 
in Fig.~\ref{fig:scatters} for comparison, 
with blue curves depicting the fitting functions and shaded areas 
representing the root-mean-square of the residual. Our results 
reproduce the non-linearity in both ends, and match the observations at all redshifts.
The presence of the non-linearity makes the linear fitting uncertain 
against the choice of the lower bound of the stellar mass, the choice of the 
likelihood function, and the galaxy sample used. In the following, we 
quantify the size-mass relations using two alternative methods that are more robust 
against these issues.

The first method is to only consider the evolution of the galaxy size 
in a narrow bin around a fixed mass. Fig.~\ref{fig:evolution} shows the 
evolution of the size-mass relation at a fixed mass of 
$M_{\rm *, bulge} = 5 \times 10^{10} \msun$. For comparison, we also show the size 
evolution obtained from various observations covering a range of redshift from
$z\sim 0$ to $z \approx 3.5$. Results obtained from different data and analyses 
are consistent with each other, with a variance of $\approx 0.2$ dex.
Our model prediction follows closely the observational trend, showing a decrease of 
$\approx 0.5$ dex in galaxy size from $z=0$ to $z=3.5$. 

The second method is to fit the size-mass relation with a non-linear function.
As discussed around equation~\eqref{eq:def-r-bulge}, the size of the bulge is
a constant fraction of the halo radius $r_{\rm f}$ in the fast assembly phase. 
Given the relation between the mass and radius of dark matter halos, the 
size-mass relation can be understood in terms of the halo mass-stellar mass 
relation that is established in the fast assembly phase. In terms of causality, 
the bulge size should be considered as the independent variable, while the bulge mass 
should be considered as a function of the bulge size. Thus, using the double-power-law
form of the stellar mass-halo mass relation found by \citet{yangEVOLUTIONGALAXYDARK2012},
we can fit the size-mass relation at a given redshift by a double-power-law 
function of the form
\begin{equation}\label{eq:fit-m-r}
    M_{\rm *, bulge} = M_{\rm 0, bulge} \frac{
        (r_{\rm bulge} / r_{\rm 0,bulge})^{\alpha_{\rm bulge} + \beta_{\rm bulge}}
    }{
        (1 +  r_{\rm bulge} / r_{\rm 0,bulge} )^{\beta_{\rm bulge}}
    } \,.
\end{equation}
The best-fit parameters, $M_{\rm 0,bulge}$, $r_{\rm 0,bulge}$, $\alpha_{\rm bulge}$, and $\beta_{\rm bulge}$, are listed 
in Table~\ref{tab:parameters}, and we show the fitting functions at 
$z \leqslant 9$ in the center panel of Fig.~\ref{fig:fitting}.
For comparison, we show the $r_{\rm v}$-$M_{\rm v}$ relation of halos,
which has a power-law index of $\beta_{\rm v} = 1/3$ at any redshift. We also show 
the $r_{\rm f}$-$M_{\rm f}$ relation obtained from a sample of simulated halos at $z=0$ 
(sample $S_{\rm h,large}$ defined in \citetalias{moTwophaseModelGalaxy2023}). This relation 
has a power-law index $\beta_{\rm f,max} = 0.41$, which is larger than 
$\beta_{\rm v}$ due to the fact that halos with higher mass have lower concentrations, 
or equivalently, later transitions from the fast assembly phase 
to the slow assembly (see \S3.3 of \citetalias{moTwophaseModelGalaxy2023}). The size-mass relations obtained 
from the fitting at different redshifts are roughly parallel to each other 
in the log-log space. At $M_{\rm *, bulge}\sim 10^{10.5} \msun$, the predicted size-mass 
relation is roughly parallel to the $r_{\rm f}$-$M_{\rm f}$ relation, with a
slope close to $\beta_{\rm f,max}$. At $M_{\rm *, bulge} < 10^{9.5} \msun$, the 
size-mass relation is flattened, with a slope even less than $\beta_{\rm v}$. 
At the massive end, $M_{\rm *, bulge} > 10^{11} \msun$, the size-mass relation shows an 
upturn, produced by the size growth during the period where the mass growth 
is stalled. The predicted increase is about $1\,$dex ($0.5\,$dex)
from $z = 9$ ($z = 3$) to $z=0$.

Since the growth of the bulge in a fast accreting halo is associated with the 
growth of its SMBH, bulge properties are expected to be tightly correlated with 
the SMBH mass, $M_{\rm bh}$. The size of a dynamically hot galaxy, $r_{\rm bulge}$, 
is thus expected to be correlated with $M_{\rm bh}$. 
To see this, we fit the $r_{\rm bulge}$-$M_{\rm bh}$ relations predicted 
by our model at different redshifts for dynamically hot galaxies 
using the same functional form as equation~\eqref{eq:fit-m-r}.
The best-fit parameters are listed in Table~\ref{tab:parameters} with a 
subscript `bh', and the fitting functions are plotted in the right panel of
Fig.~\ref{fig:fitting}.
The $r_{\rm bulge}$-$M_{\rm bh}$
relation shows similar behavior to the $r_{\rm bulge}$-$M_{\rm *,bulge}$
relation at low redshift. 
This is expected because of the $M_{\rm bh}$-$M_{\rm *,bulge}$ relation 
observed \citep[e.g.]{greeneIntermediateMassBlackHoles2020,
grahamAppreciatingMergersUnderstanding2023,
zhuangEvolutionaryPathsActive2023} at $z\approx 0$ 
and reproduced by our model (see Fig.~7 of \citetalias{moTwophaseModelGalaxy2023}).
The observed $M_{\rm bh}$-$M_{\rm *,bulge}$ relation has a power-law 
index about $3/2$ for early-type galaxies with $M_{\rm *,bulge} \gtrsim 10^{10} \msun$. 
This, combined with the power-law indices $\beta_{\rm v}$ and 
$\beta_{\rm f,max}$, gives the solid and dashed black lines
in the right penal of Fig.~\ref{fig:fitting}, respectively. 
Our predicted $r_{\rm bulge}$-$M_{\rm bh}$ relation for $M_{\rm bh} \gtrsim 10^{7}\msun$ 
at $z \approx 0$ has the expected power-law behavior. At higher redshift, 
$M_{\rm bh}$ is shifted leftward relative to $M_{\rm *, bulge}$, because 
low-mass SMBHs have not reached the self-regulation regime and do not 
follow the $M_{\rm bh}$-$M_{\rm *,bulge}$ scaling relation.

For reference, we also show the bulge size-halo mass ($M_{\rm f}$) relations 
for dynamically hot galaxies at different redshifts in the left panel 
of Fig.~\ref{fig:fitting}. These relations are obtained by fitting the 
model predictions at individual redshifts with a power-law function of the form
\begin{equation}
    \log r_{\rm bulge} = \left( \frac{M_{\rm f}}{M_{\rm 0,f}} \right)^{\beta_{\rm f}},
\end{equation}
and the best-fit parameters are also listed in Table~\ref{tab:parameters}. 
At all redshifts, the power-law index $\beta_{\rm f}$ is contained in a 
narrow range from $0.33$ to $0.37$. 
At a given $M_{\rm f}$, the bulge size increases as redshift decreases due 
to the decrease of the density threshold, approximately $200$ times the mean 
density of the Universe, required for the collapse of CDM halos.
In the following, we use these results to interpret 
the physical origin of the size-mass relation of dynamically hot galaxies.

\begin{center}
    \begin{table*} 
    \caption{
        Best-fit parameters of the bulge size
        ($r_{\rm bulge}$)-halo mass ($M_{\rm f}$) relation,
        bulge size-bulge mass ($M_{\rm *, bulge}$) relation,
        and bulge size-SMBH mass ($M_{\rm bh}$) relation
        of dynamically hot galaxies predicted by our model at different redshift ($z$).
        See Fig.~\ref{fig:fitting} for a plot of fitted functions and 
        \S\ref{ssec:compare-with-obs} for a detailed description.
        No observational data for $r_{\rm bulge}$ are currently available at $z \gtrsim 4$.
        The root-mean-square error, $\sigma_{\log r_{\rm bulge}}$, is listed 
        reference.
    }
    \scriptsize
    \begin{tabularx}{0.98\textwidth}{c || c c c || c c c c c || c c c c c}
        \hline
            $z$ 		 
            &
            $\log\,M_{\rm 0,f}$
            &
            $\beta_{\rm f}$
            &
            $\sigma_{\log\,r_{\rm bulge}}$
            &
            $\log\,r_{\rm 0,bulge}$
            &
            $\log\,M_{\rm 0,bulge}$
            &
            $\alpha_{\rm bulge}$
            &
            $\beta_{\rm bulge}$
            &
            $\sigma_{\log\,r_{\rm bulge}}$
            &
            $\log\,r_{\rm 0,bh}$
            &
            $\log\,M_{\rm 0,bh}$
            &
            $\alpha_{\rm bh}$
            &
            $\beta_{\rm bh}$
            &
            $\sigma_{\log\,r_{\rm bulge}}$
        \\ 		 
            & $[\msun]$ & & $[{\rm dex}]$
            &  $[\kpc]$ & $[\msun]$ & & & $[{\rm dex}]$
            & $[\kpc]$ & $[\msun]$ & & & $[{\rm dex}]$ \\
        \hline
        \hline
        0.1    &   11.6 & 0.364 & 0.226 &  0.515 &  12.34 & -0.716 &  4.62 & 0.331  &  0.565 &  11.2 & -1.86  & 8.96  & 0.31 \\
        0.75   &  11.93 & 0.367 & 0.227 &  0.425 &  12.33 & -0.496 &  4.29 & 0.333  &  0.455 & 11.87 & -3.11  & 10.6   & 0.4 \\
        1.25   &  12.14 & 0.371 &  0.23 &  0.515 &  12.59 & -0.652 &  4.12 & 0.326  &  0.334 & 12.24 & -4.05  & 12.1 & 0.479 \\
        1.75   &  12.32 & 0.359 & 0.219 &  0.696 &  13.09 &  -1.33 &  4.53 & 0.304  &  0.224 & 12.39 & -4.73    & 13 & 0.523 \\
        2.5    &  12.54 & 0.356 & 0.208 &  0.867 &   13.8 &  -2.96 &  6.07 & 0.297  &  0.153 & 12.74 & -6.32  & 14.7 & 0.567 \\
        3.25   &   12.8 & 0.354 & 0.205 &  0.807 &  14.19 &  -4.24 &   7.5 & 0.326  & 0.0528 & 12.84 & -7.44  & 15.7 & 0.566 \\
        4      &  12.96 & 0.349 & 0.201 &  0.656 &  14.32 &  -4.64 &  8.13 & 0.362  & 0.00251 & 12.85 &  -8.5  & 16.6 & 0.556 \\
        5      &  13.21 & 0.343 & 0.208 &  0.425 &  14.33 &  -4.61 &  8.43 & 0.411  & -0.0477 & 12.21 & -9.06  & 16.3 & 0.537 \\
        7      &  13.52 & 0.347 &   0.2 & 0.0226 &  14.06 &  -3.81 &  8.15 & 0.476  & -0.249 & 8.629 & -5.09  & 9.18 & 0.492 \\
        9      &  13.87 & 0.333 & 0.209 & -0.299 &  13.43 &  -2.58 &   7.1 & 0.488  &   -0.5 & 5.258 & -1.37  & 2.68  & 0.56 \\
        \hline
    \end{tabularx}
    \label{tab:parameters}
    \end{table*}
    \end{center}

\subsection{The physical origin of the size-mass relation}
\label{ssec:interp-size-mass}

\begin{figure} \centering
    \includegraphics[width=\columnwidth]{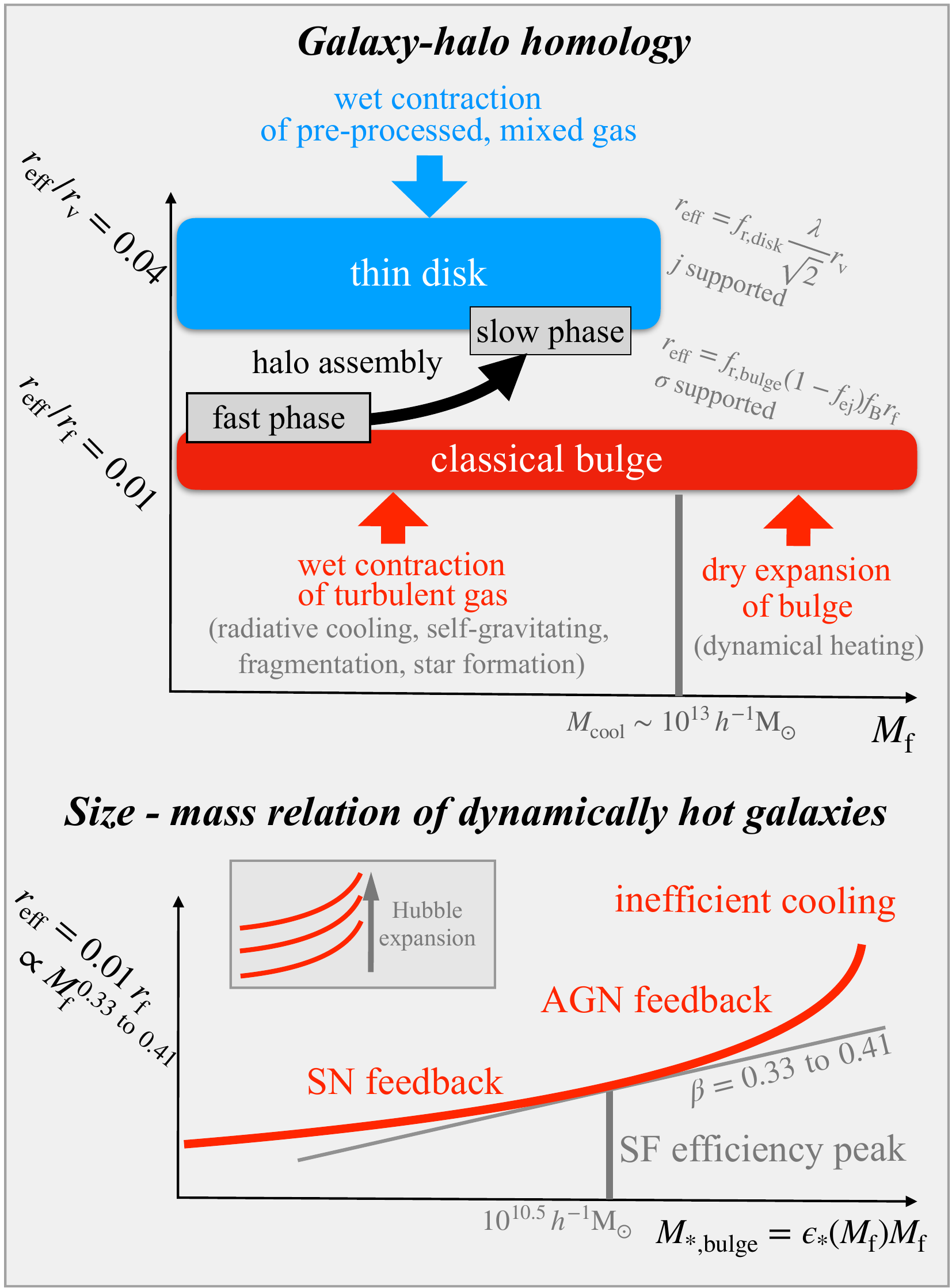}
    \caption{
        A schematic figure showing the physical origin of galaxy-halo 
        homology ({\bf top panel}, see \S\ref{sec:model} for the details) 
        and the size-mass relation
        of dynamically hot galaxies ({\bf bottom panel}, see \S\ref{sec:interp-obs}
        for the details). In both panels, we show the properties in both axes 
        in a logarithmic scale. This figure demonstrates that the galaxy-halo 
        homology (i.e. constant ratio between the galaxy size and halo size) 
        of dynamically hot component is built up by the {\em wet contraction} 
        of turbulent gas, and preserved by the {\em dry expansion} 
        of the stellar bulge. Consequently, the size-mass relation of dynamically
        hot galaxies at a given redshift directly reflects the stellar 
        mass-halo mass relation,  which is shaped by 
        {\em cooling and feedback processes}, and the evolution of the 
        size-mass relation originates from the evolution of the
        halo size-halo mass relation.
    }
    \label{fig:size-mass-schematic}
\end{figure}

Given that the observed features in the size-mass relation are
well reproduced by our model, we can use our model to understand 
the physical origin of these features. The basic underpinning is a simple combination 
of the homology relation, $r_{\rm f} \approx 100\, r_{\rm bulge}$, and the 
stellar mass-halo mass relation (SMHMR), at any given redshift $z$, 
as described below. 
\begin{enumerate}[topsep=0pt]
\item 
At $M_{\rm *, bulge}\approx 10^{10.5}\msun$ ($M_{\rm f}\approx 10^{12} \msun$), 
the star formation efficiency (described by $M_{\rm *, bulge}/M_{\rm f}$) 
reaches its peak, which signals the regime where AGN feedback is about to become 
effective. Consequently, the growth of the stellar mass follows the growth of
the halo mass (i.e. ${\rm d}\log M_{\rm *, bulge}/{\rm d}\log M_{\rm f}$ = 1), 
and the growth of the bulge size follows the growth of the halo size 
due to the wet contraction. The size-mass relation thus has a power-law index 
close to $\beta_{\rm f}$, the power index of the size-mass relation for halos in 
fast assembly. 
Note that dynamically hot galaxies ($f_{\rm bulge}>0.75$) have 
halo transition redshift, $z_{\rm f}$, close to the redshift $z$ in question,
and hence $M_{\rm f} \approx M_{\rm v}$, $M_{\rm *,bulge}\approx M_*$. 
The shape of the $M_{\rm *,bulge}$-$M_{\rm f}$ relation of these 
galaxies (see Appendix~\ref{sec:app:sfe}) thus follows that of their
$M_{\rm *}$-$M_{\rm v}$ relation \citep[e.g.][]{yangConstrainingGalaxyFormation2003, 
yangEVOLUTIONGALAXYDARK2012,behrooziAVERAGESTARFORMATION2013}.
\item 
At $M_{\rm *, bulge} < 10^{9.5} \msun$, the star formation efficiency is suppressed by SN feedback, 
which produces a lower $M_{\rm *, bulge}$ for a given $M_{\rm f}$ than expected 
from the peak of the star formation efficiency. However, the size of the bulge is still controlled 
by the wet contraction, and thus by $r_{\rm f}$ according to the homology relation.
This produces a flattening in the size-mass relation, where the power-law index is smaller 
than $\beta_{\rm f}$, or even smaller than $\beta_{\rm v}$, depending on the star formation efficiency.
\item 
At $M_{\rm *, bulge} > 10^{10.5} \msun$, the star formation efficiency can be reduced 
significantly by AGN feedback, again leading to a lower $M_{\rm *, bulge}$ at 
given $M_{\rm f}$ than that expected from the star formation at the peak. 
The galaxy-halo homology is again fixed by the wet contraction, which produces 
a steepening in the size-mass relation with a power index larger than $\beta_{\rm f}$.
\item 
In the massive end, $M_{\rm *, bulge} > 10^{11} \msun$, or $M_{\rm f} > 10^{13}\msun$, 
star formation is quenched due to ineffective cooling, producing a dry system 
with fixed stellar mass. However, the homologous galaxy-halo size relation is 
still preserved by dynamical heating driven by the fast halo assembly. 
The synergy of the fixed stellar mass and the expansion of the bulge results 
in a significant upturn in the high-mass end of the size-mass relation.
\item 
At a given bulge mass, the increase of the bulge size with decreasing redshift 
is produced by the increase of $r_{\rm f}$ due to the decrease of the density
threshold required for the collapse of CDM halos.
\end{enumerate}

The above interpretations, including the physical origin of the galaxy-halo 
homology, the non-linear transformation due to the incorporation of the SMHMR,
and the time evolution due to the Hubble expansion, are summarized 
schematically in Fig.~\ref{fig:size-mass-schematic}. This figure 
also includes the formation scenario of disk galaxies presented 
in \citet{moFormationGalacticDiscs1998}. A mixed system of bulge and disk
may then be considered as a combination of the two components, with the 
mass ratio between the two determined by the transition time of the halo.
The effective radius of such a system is expected to fall between 
that of the dynamically hot galaxy (classical bulge) and 
dynamically cold galaxy (thin disk), 
i.e. between $0.01\, r_{\rm f}$  and $0.04\, r_{\rm v}$, 
which is consistent with the scaling relations found in observations 
with abundance matching \citep{kravtsovSizeVirialRadiusRelation2013,
huangRelationsSizesGalaxies2017,somervilleRelationshipGalaxyDark2018} 
and gravitational weak lensing \citep{mishraStellarMassDependence2023}.

\section{Summary and Discussion}
\label{sec:summary}

We use the two-phase model developed in \citetalias{moTwophaseModelGalaxy2023} to predict the size-mass
relation for dynamically hot galaxies (such as ellipticals and bulges). We 
find a remarkable agreement between our model predictions and the observed 
size-mass relation up to $z \approx 4$, with all observed features well reproduced.
We provide a clear and physically-motivated interpretation regarding the origin of the
size-mass relation. Our analyses and results are summarized as follows.
\begin{enumerate}
\item {\em Two-phase assembly of halos}: 
    Our model is based on the fact that the growth of cold dark matter halos 
    consists of an early fast phase of assembly followed by a slow assembly phase. 
\item {\em Build-up of the galaxy-halo homology}: 
    In the fast assembly phase, 
    rapid radiative cooling causes rapid collapse of the halo gas, 
    producing a turbulent gas cloud. The contraction of the gas 
    is halted once the cloud becomes self-gravitation and fragments to form 
    dense sub-clouds that can move ballistically.  
    Stars that form in sub-clouds inherit their spatial distribution and dynamical hotness,
    producing a stellar bulge with a characteristic size ($r_{\rm bulge}$) 
    proportional to the halo virial radius ($r_{\rm f}$). Detailed modeling and 
    calibration give a homologous relation, $r_{\rm f} \approx\, 100\,r_{\rm bulge}$, 
    independent of the halo mass and redshift.
\item {\em Preservation of the galaxy-halo homology}: 
    For a massive halo with mass exceeding the cooling threshold
    in the fast phase, the gas supply is cut off and the stellar bulge becomes
    a dry system in the regime of quenching.
    The dry expansion caused by dynamical `heating' associated with the fast halo assembly  
    preserves the galaxy-halo homology, extending the halo size-bulge size 
    relation to the end of the fast assembly phase.
\item {\em Freeze of the bulge growth}: As a halo enters the slow assembly phase,
    the reduced gas fraction and the longer time for cooling provide the condition 
    for the formation of a dynamically cold disk, thereby terminating the mass growth 
    of the bulge. The stable gravitational potential also reduces dynamical heating, 
    which freezes the growth of the bulge size. The bulge mass and size are then 
    preserved during the entire phase of the slow assembly.
\item {\em The size-mass relation of dynamically hot galaxies}:
    The combination of the galaxy-halo homology relation and the 
    stellar mass-halo mass relation is the key to producing the size-mass relation. 
    The non-linearity in the size-mass relation is a
    direct consequence of the star formation efficiency ($M_{\rm *, bulge}/M_{\rm f}$) in 
    halos, which is controlled by gas cooling and feedback, 
    and the time evolution of the size-mass relation is a direct reflection
    of the halo radius at a given halo mass, which increases with time 
    due to Hubble expansion.
\item {\em Comparison with observations}:
    A comparison of our model prediction with observations at $z \lesssim 4$ indicates 
    that the model is able to reproduce the observed size-mass relation, 
    including its slope at $M_{\rm *, bulge}>10^{10}\msun$, the 
    non-linear flattening in the low-mass end, the non-linear upturn in the high-mass end,
    and the time evolution of the relation (Figs.~\ref{fig:scatters} and \ref{fig:evolution}).
\item {\em Prediction for future observations}:
    Our model makes predictions for the 
    size ($r_{\rm bulge}$)-halo mass ($M_{\rm f}$) relation,
    size-bulge mass ($M_{\rm *,bulge}$) relation, 
    and size-SMBH mass ($M_{\rm bh}$) relation,
    of dynamically hot galaxies at $z \gtrsim 4$ (Fig.~\ref{fig:fitting}),
    which can be tested by future observations.
\end{enumerate}

The galaxy-halo homology relation of dynamically hot galaxies, 
$r_{\rm f} \approx 100\, r_{\rm bulge}$, can be tested using weak 
gravitational lensing observations. Investigations 
in this direction have been conducted \citep[e.g.][]{mishraStellarMassDependence2023},
and a nearly constant $r_{\rm bulge}$-$r_{\rm v}$ ratio
has been identified, albeit with some subtle dependence on 
the stellar mass. Our analysis suggests that the correlation should
be tested on central elliptical galaxies, and the halo property to be 
inferred should be $r_{\rm f} \approx 4 r_{\rm s}$, where $r_{\rm s}$
is the characteristic radius of the NFW profile.
The predicted constant $r_{\rm bulge}$-$r_{\rm f}$ ratio, 
with little secondary dependence on other properties, such as 
stellar mass, halo mass and redshift, can then be checked by observations. 

Our model for the formation of dynamically hot central galaxies suggests 
that the galaxy-halo connection may be more properly 
described by $M_*$ and $M_{\rm f}$, rather than by $M_*$ and $M_{\rm v}$, 
as the use of $M_{\rm f}$ provides a more precise prediction of $M_*$ and 
avoids the uncertainties in the definition of halo boundary. The homology 
relation of dynamically hot galaxies also suggests that the bulge size
may also contain information about the halo mass. Thus, the synergistic use of galaxy 
size and stellar mass may provide a more accurate estimate of the 
halo mass, which can be calibrated and applied to observations. 
The predicted tight relation between the bulge size 
and the SMBH mass suggests that this relation should be analyzed   
to provide additional constraints on SMBH mass in observations.

Our model includes only central galaxies, thus missing  
environmental effects specific for satellite galaxies. However, as the bulge is not as 
extended as the disk, tidal stripping is expected to have only minor  
effects on the structure of the bulge. Observations by \citet{guoStructuralPropertiesCentral2009} 
from the SDSS support this expectation. Our model does not explicitly 
incorporate mergers in the growth of the bulge. Mergers are expected to be important  
for the formation of massive ellipticals. The effect of minor mergers on 
the mass growth is expected to be as small as $\approx 10\%$, as indicated by 
numerical simulations \citep[e.g.][]{hopkinsMergersBulgeFormation2010} and 
semi-analytic models \citep[e.g.][]{lotzMajorMinorGalaxy2011}. Furthermore, 
their effects on the size growth have already been included in the dry 
expansion. For major mergers, particularly those involving gas-rich progenitors, 
there are debates as to whether the in-situ star formation in the merger remnant 
or the accretion of ex-situ stars dominates the bulge growth. The in-situ mass and 
size growths have been included in our model for the star formation in the SGC. 
The ex-situ accretion from major mergers yields approximately a 1:1 mass-size growth, 
thus is degenerate with the in-situ growth. Consequently, the inclusion of mergers 
is not expected to change the main conclusion of our model for the size-mass relation of 
dynamically hot galaxies. Numerical simulations have been used to 
disentangle the in-situ versus ex-situ contributions \citep[e.g.][]{oserTwoPhasesGalaxy2010}. 
However, current simulations are yet to predict sizes for elliptical 
galaxies to match observational data \citep{mishraStellarMassDependence2023},  
and need improvements before reaching any definitive conclusions. 
Future observations of galaxies in the high-$z$ Universe, where 
ex-situ accretion may play a less important role, may help to resolve
the degeneracy \citep[e.g.][]{jiReconstructingAssemblyMassive2022,
jiReconstructingAssemblyMassive2023}. In addition, detailed analyses of the 
profiles of local galaxies using spatially resolved spectroscopy and stellar population 
synthesis can also shed light on the problem \citep[e.g.][]{liPMaNGAGradientsRecent2015,
linSDSSIVMaNGAInsideout2019,bluckAreGalacticStar2020}.

\section*{Acknowledgements}

YC is funded by the China Postdoctoral Science Foundation (grant No. 2022TQ0329).
HJM thanks T.D. Lee Institute and Shanghai Jiaotong University for hosting his 
sabbatical during which the work was done.
This work is also supported by the National Natural Science Foundation of China 
(NSFC, Nos. 12192224, 11733004 and 11890693) and CAS Project for Young Scientists 
in Basic Research (grant No. YSBR-062).
YC thanks Kai Wang, Yu Rong, Wentao Luo, Enci Wang, Hao Li and Hui Hong for 
their valuable insights and discussions. 
The authors would like to express their gratitude to the Tsinghua Astrophysics 
High-Performance Computing platform at Tsinghua University and the 
Supercomputer Center of the University of Science and Technology of China for 
providing the necessary computational and data storage resources that have 
significantly contributed to the research results presented in this paper.
The computations and presentations in this paper are supported by various software 
tools, including the HPC toolkits 
\softwarenamestyle[Hipp] \citep{chenHIPPHIghPerformancePackage2023}\footnote{\url{https://github.com/ChenYangyao/hipp}}
and \softwarenamestyle[PyHipp]\footnote{\url{https://github.com/ChenYangyao/pyhipp}},
interactive computation environment 
\softwarenamestyle[IPython] \citep{perezIPythonSystemInteractive2007},
numerical libraries \softwarenamestyle[NumPy] \citep{harrisArrayProgrammingNumPy2020}, 
\softwarenamestyle[Astropy] \citep{
robitailleAstropyCommunityPython2013,
astropycollaborationAstropyProjectBuilding2018,
astropycollaborationAstropyProjectSustaining2022}
and \softwarenamestyle[SciPy] \citep{virtanenSciPyFundamentalAlgorithms2020}, 
as well as the graphical library 
\softwarenamestyle[Matplotlib] \citep{hunterMatplotlib2DGraphics2007}. 
This research has made extensive use of the arXiv and NASA’s Astrophysics Data System.
Data compilations used in this paper have been made much more accurate and 
efficient by the software \softwarenamestyle[WebPlotDigitizer]. 

\section*{Data Availability}
\label{sec:data-availability}
 
The code repository \softwarenamestyle[TwoPhaseGalaxyModel]\footnote{\url{https://github.com/ChenYangyao/two-phase-galaxy-model}}
implements the model described \citetalias{moTwophaseModelGalaxy2023}. 
All data used in this paper, including data points displayed in figures
and observational results used for calibration and comparison, 
will be distributed along with the repository.


\bibliographystyle{mnras}
\bibliography{references}

\begin{thebibliography}{}
\makeatletter
\relax
\def\mn@urlcharsother{\let\do\@makeother \do\$\do\&\do\#\do\^\do\_\do\%\do\~}
\def\mn@doi{\begingroup\mn@urlcharsother \@ifnextchar [ {\mn@doi@}
  {\mn@doi@[]}}
\def\mn@doi@[#1]#2{\def\@tempa{#1}\ifx\@tempa\@empty \href
  {http://dx.doi.org/#2} {doi:#2}\else \href {http://dx.doi.org/#2} {#1}\fi
  \endgroup}
\def\mn@eprint#1#2{\mn@eprint@#1:#2::\@nil}
\def\mn@eprint@arXiv#1{\href {http://arxiv.org/abs/#1} {{\tt arXiv:#1}}}
\def\mn@eprint@dblp#1{\href {http://dblp.uni-trier.de/rec/bibtex/#1.xml}
  {dblp:#1}}
\def\mn@eprint@#1:#2:#3:#4\@nil{\def\@tempa {#1}\def\@tempb {#2}\def\@tempc
  {#3}\ifx \@tempc \@empty \let \@tempc \@tempb \let \@tempb \@tempa \fi \ifx
  \@tempb \@empty \def\@tempb {arXiv}\fi \@ifundefined
  {mn@eprint@\@tempb}{\@tempb:\@tempc}{\expandafter \expandafter \csname
  mn@eprint@\@tempb\endcsname \expandafter{\@tempc}}}

\bibitem[\protect\citeauthoryear{{Astropy Collaboration} et~al.,}{{Astropy
  Collaboration} et~al.}{2018}]{astropycollaborationAstropyProjectBuilding2018}
{Astropy Collaboration} et~al., 2018, \mn@doi [The Astronomical Journal]
  {10.3847/1538-3881/aabc4f}, 156, 123

\bibitem[\protect\citeauthoryear{{Astropy Collaboration} et~al.,}{{Astropy
  Collaboration}
  et~al.}{2022}]{astropycollaborationAstropyProjectSustaining2022}
{Astropy Collaboration} et~al., 2022, \mn@doi [The Astrophysical Journal]
  {10.3847/1538-4357/ac7c74}, 935, 167

\bibitem[\protect\citeauthoryear{Behroozi, Wechsler  \& Conroy}{Behroozi
  et~al.}{2013}]{behrooziAVERAGESTARFORMATION2013}
Behroozi P.~S.,  Wechsler R.~H.,   Conroy C.,  2013, \mn@doi [ApJ]
  {10.1088/0004-637X/770/1/57}, 770, 57

\bibitem[\protect\citeauthoryear{Bernardi, Meert, Vikram, {Huertas-Company},
  Mei, Shankar  \& Sheth}{Bernardi
  et~al.}{2012}]{bernardiSystematicEffectsSizeluminosity2012}
Bernardi M.,  Meert A.,  Vikram V.,  {Huertas-Company} M.,  Mei S.,  Shankar
  F.,   Sheth R.~K.,  2012, Systematic Effects on the Size-Luminosity Relation:
  Dependence on Model Fitting and Morphology, \mn@doi{10.48550/arXiv.1211.6122}

\bibitem[\protect\citeauthoryear{Bett, Eke, Frenk, Jenkins, Helly  \&
  Navarro}{Bett et~al.}{2007}]{bettSpinShapeDark2007}
Bett P.,  Eke V.,  Frenk C.~S.,  Jenkins A.,  Helly J.,   Navarro J.,  2007,
  \mn@doi [Monthly Notices of the Royal Astronomical Society]
  {10.1111/j.1365-2966.2007.11432.x}, 376, 215

\bibitem[\protect\citeauthoryear{Bezanson, {van Dokkum}, Tal, Marchesini,
  Kriek, Franx  \& Coppi}{Bezanson
  et~al.}{2009}]{bezansonRelationCompactQuiescent2009}
Bezanson R.,  {van Dokkum} P.~G.,  Tal T.,  Marchesini D.,  Kriek M.,  Franx
  M.,   Coppi P.,  2009, \mn@doi [The Astrophysical Journal]
  {10.1088/0004-637X/697/2/1290}, 697, 1290

\bibitem[\protect\citeauthoryear{Blank, Macci{\`o}, Dutton  \& Obreja}{Blank
  et~al.}{2019}]{blankNIHAOXXIIIntroducing2019}
Blank M.,  Macci{\`o} A.~V.,  Dutton A.~A.,   Obreja A.,  2019, \mn@doi
  [Monthly Notices of the Royal Astronomical Society] {10.1093/mnras/stz1688},
  487, 5476

\bibitem[\protect\citeauthoryear{Bluck, Maiolino, S{\'a}nchez, Ellison, Thorp,
  Piotrowska, Teimoorinia  \& Bundy}{Bluck
  et~al.}{2020}]{bluckAreGalacticStar2020}
Bluck A. F.~L.,  Maiolino R.,  S{\'a}nchez S.~F.,  Ellison S.~L.,  Thorp M.~D.,
   Piotrowska J.~M.,  Teimoorinia H.,   Bundy K.~A.,  2020, \mn@doi [Monthly
  Notices of the Royal Astronomical Society] {10.1093/mnras/stz3264}, 492, 96

\bibitem[\protect\citeauthoryear{Bullock, Dekel, Kolatt, Kravtsov, Klypin,
  Porciani  \& Primack}{Bullock
  et~al.}{2001}]{bullockUniversalAngularMomentum2001}
Bullock J.~S.,  Dekel A.,  Kolatt T.~S.,  Kravtsov A.~V.,  Klypin A.~A.,
  Porciani C.,   Primack J.~R.,  2001, \mn@doi [ApJ] {10.1086/321477}, 555, 240

\bibitem[\protect\citeauthoryear{Burkert et~al.,}{Burkert
  et~al.}{2016}]{burkertANGULARMOMENTUMDISTRIBUTION2016}
Burkert A.,  et~al., 2016, \mn@doi [ApJ] {10.3847/0004-637X/826/2/214}, 826,
  214

\bibitem[\protect\citeauthoryear{Chen \& Wang}{Chen \&
  Wang}{2023}]{chenHIPPHIghPerformancePackage2023}
Chen Y.,  Wang K.,  2023, Astrophysics Source Code Library, p. ascl:2301.030

\bibitem[\protect\citeauthoryear{Ciotti, Lanzoni  \& Volonteri}{Ciotti
  et~al.}{2007}]{ciottiImportanceDryWet2007}
Ciotti L.,  Lanzoni B.,   Volonteri M.,  2007, \mn@doi [The Astrophysical
  Journal] {10.1086/510773}, 658, 65

\bibitem[\protect\citeauthoryear{Conroy}{Conroy}{2013}]{conroyModelingPanchromaticSpectral2013}
Conroy C.,  2013, \mn@doi [Annual Review of Astronomy and Astrophysics]
  {10.1146/annurev-astro-082812-141017}, 51, 393

\bibitem[\protect\citeauthoryear{Cui, Dav{\'e}, Peacock,
  {Angl{\'e}s-Alc{\'a}zar}  \& Yang}{Cui
  et~al.}{2021}]{cuiOriginGalaxyColour2021}
Cui W.,  Dav{\'e} R.,  Peacock J.~A.,  {Angl{\'e}s-Alc{\'a}zar} D.,   Yang X.,
  2021, The Origin of Galaxy Colour Bimodality in the Scatter of the
  {{Stellar-to-Halo Mass Relation}} (\mn@eprint {arxiv} {2105.12145}),
  \mn@doi{10.48550/arXiv.2105.12145}

\bibitem[\protect\citeauthoryear{Dav{\'e}, {Angl{\'e}s-Alc{\'a}zar}, Narayanan,
  Li, Rafieferantsoa  \& Appleby}{Dav{\'e}
  et~al.}{2019}]{daveSimbaCosmologicalSimulations2019}
Dav{\'e} R.,  {Angl{\'e}s-Alc{\'a}zar} D.,  Narayanan D.,  Li Q.,
  Rafieferantsoa M.~H.,   Appleby S.,  2019, \mn@doi [Monthly Notices of the
  Royal Astronomical Society] {10.1093/mnras/stz937}, 486, 2827

\bibitem[\protect\citeauthoryear{Deng, Li, Kannan, Smith, Vogelsberger  \&
  Bryan}{Deng et~al.}{2024}]{dengSimulatingIonizationFeedback2024}
Deng Y.,  Li H.,  Kannan R.,  Smith A.,  Vogelsberger M.,   Bryan G.~L.,  2024,
  \mn@doi [Monthly Notices of the Royal Astronomical Society]
  {10.1093/mnras/stad3202}, 527, 478

\bibitem[\protect\citeauthoryear{Desmond \& Wechsler}{Desmond \&
  Wechsler}{2015}]{desmondTullyFisherMasssizeRelations2015}
Desmond H.,  Wechsler R.~H.,  2015, \mn@doi [Monthly Notices of the Royal
  Astronomical Society] {10.1093/mnras/stv1978}, 454, 322

\bibitem[\protect\citeauthoryear{Fan, Lapi, De~Zotti  \& Danese}{Fan
  et~al.}{2008}]{fanDramaticSizeEvolution2008}
Fan L.,  Lapi A.,  De~Zotti G.,   Danese L.,  2008, \mn@doi [The Astrophysical
  Journal] {10.1086/595784}, 689, L101

\bibitem[\protect\citeauthoryear{Graham}{Graham}{2013}]{grahamEllipticalDiskGalaxy2013}
Graham A.~W.,  2013, in Oswalt T.~D.,  Keel W.~C.,  eds, , Planets, Stars and
  Stellar Systems: {{Volume}} 6: {{Extragalactic}} Astronomy and Cosmology.
Springer Netherlands, Dordrecht, pp 91--139,
  \mn@doi{10.1007/978-94-007-5609-0_2}

\bibitem[\protect\citeauthoryear{Graham \& Sahu}{Graham \&
  Sahu}{2023}]{grahamAppreciatingMergersUnderstanding2023}
Graham A.~W.,  Sahu N.,  2023, \mn@doi [Monthly Notices of the Royal
  Astronomical Society] {10.1093/mnras/stac2019}, 518, 2177

\bibitem[\protect\citeauthoryear{Greene, Strader  \& Ho}{Greene
  et~al.}{2020}]{greeneIntermediateMassBlackHoles2020}
Greene J.~E.,  Strader J.,   Ho L.~C.,  2020, \mn@doi [Annual Review of
  Astronomy and Astrophysics] {10.1146/annurev-astro-032620-021835}, 58, 257

\bibitem[\protect\citeauthoryear{Guo et~al.,}{Guo
  et~al.}{2009}]{guoStructuralPropertiesCentral2009}
Guo Y.,  et~al., 2009, \mn@doi [Monthly Notices of the Royal Astronomical
  Society] {10.1111/j.1365-2966.2009.15223.x}, 398, 1129

\bibitem[\protect\citeauthoryear{Harris et~al.,}{Harris
  et~al.}{2020}]{harrisArrayProgrammingNumPy2020}
Harris C.~R.,  et~al., 2020, \mn@doi [Nature] {10.1038/s41586-020-2649-2}, 585,
  357

\bibitem[\protect\citeauthoryear{Hearin, {Chaves-Montero}, Becker  \&
  Alarcon}{Hearin et~al.}{2021}]{hearinDifferentiableModelAssembly2021}
Hearin A.~P.,  {Chaves-Montero} J.,  Becker M.~R.,   Alarcon A.,  2021, \mn@doi
  [The Open Journal of Astrophysics] {10.21105/astro.2105.05859}, 4, 7

\bibitem[\protect\citeauthoryear{Hopkins, Bundy, Hernquist, Wuyts  \&
  Cox}{Hopkins et~al.}{2010a}]{hopkinsDiscriminatingPhysicalProcesses2010}
Hopkins P.~F.,  Bundy K.,  Hernquist L.,  Wuyts S.,   Cox T.~J.,  2010a,
  \mn@doi [Monthly Notices of the Royal Astronomical Society]
  {10.1111/j.1365-2966.2009.15699.x}, 401, 1099

\bibitem[\protect\citeauthoryear{Hopkins et~al.,}{Hopkins
  et~al.}{2010b}]{hopkinsMergersBulgeFormation2010}
Hopkins P.~F.,  et~al., 2010b, \mn@doi [The Astrophysical Journal]
  {10.1088/0004-637X/715/1/202}, 715, 202

\bibitem[\protect\citeauthoryear{Hopkins et~al.,}{Hopkins
  et~al.}{2018}]{hopkinsFIRE2SimulationsPhysics2018}
Hopkins P.~F.,  et~al., 2018, \mn@doi [Monthly Notices of the Royal
  Astronomical Society] {10.1093/mnras/sty1690}, 480, 800

\bibitem[\protect\citeauthoryear{Hopkins et~al.,}{Hopkins
  et~al.}{2023a}]{hopkinsFIRE3UpdatedStellar2023}
Hopkins P.~F.,  et~al., 2023a, \mn@doi [Monthly Notices of the Royal
  Astronomical Society] {10.1093/mnras/stac3489}, 519, 3154

\bibitem[\protect\citeauthoryear{Hopkins et~al.,}{Hopkins
  et~al.}{2023b}]{hopkinsWhatCausesFormation2023}
Hopkins P.~F.,  et~al., 2023b, \mn@doi [Monthly Notices of the Royal
  Astronomical Society] {10.1093/mnras/stad1902}, 525, 2241

\bibitem[\protect\citeauthoryear{Huang et~al.,}{Huang
  et~al.}{2017}]{huangRelationsSizesGalaxies2017}
Huang K.-H.,  et~al., 2017, \mn@doi [The Astrophysical Journal]
  {10.3847/1538-4357/aa62a6}, 838, 6

\bibitem[\protect\citeauthoryear{Hunter}{Hunter}{2007}]{hunterMatplotlib2DGraphics2007}
Hunter J.~D.,  2007, \mn@doi [Computing in Science \& Engineering]
  {10.1109/MCSE.2007.55}, 9, 90

\bibitem[\protect\citeauthoryear{Ito et~al.,}{Ito
  et~al.}{2023}]{itoSizeStellarMass2023}
Ito K.,  et~al., 2023, Size - {{Stellar Mass Relation}} and {{Morphology}} of
  {{Quiescent Galaxies}} at \$z{\textbackslash}geq3\$ in {{Public}}
  \${{JWST}}\$ {{Fields}}, \mn@doi{10.48550/arXiv.2307.06994}

\bibitem[\protect\citeauthoryear{Ji \& Giavalisco}{Ji \&
  Giavalisco}{2022}]{jiReconstructingAssemblyMassive2022}
Ji Z.,  Giavalisco M.,  2022, \mn@doi [ApJ] {10.3847/1538-4357/ac7f43}, 935,
  120

\bibitem[\protect\citeauthoryear{Ji \& Giavalisco}{Ji \&
  Giavalisco}{2023}]{jiReconstructingAssemblyMassive2023}
Ji Z.,  Giavalisco M.,  2023, \mn@doi [ApJ] {10.3847/1538-4357/aca807}, 943, 54

\bibitem[\protect\citeauthoryear{Kormendy \& Kennicutt}{Kormendy \&
  Kennicutt}{2004}]{kormendySecularEvolutionFormation2004}
Kormendy J.,  Kennicutt Jr. R.~C.,  2004, \mn@doi [Annual Review of Astronomy
  and Astrophysics] {10.1146/annurev.astro.42.053102.134024}, 42, 603

\bibitem[\protect\citeauthoryear{Kravtsov}{Kravtsov}{2013}]{kravtsovSizeVirialRadiusRelation2013}
Kravtsov A.~V.,  2013, \mn@doi [The Astrophysical Journal]
  {10.1088/2041-8205/764/2/L31}, 764, L31

\bibitem[\protect\citeauthoryear{Lange et~al.,}{Lange
  et~al.}{2015}]{langeGalaxyMassAssembly2015}
Lange R.,  et~al., 2015, \mn@doi [Monthly Notices of the Royal Astronomical
  Society] {10.1093/mnras/stu2467}, 447, 2603

\bibitem[\protect\citeauthoryear{Li et~al.,}{Li
  et~al.}{2015}]{liPMaNGAGradientsRecent2015}
Li C.,  et~al., 2015, \mn@doi [The Astrophysical Journal]
  {10.1088/0004-637X/804/2/125}, 804, 125

\bibitem[\protect\citeauthoryear{Li, Vogelsberger, Marinacci, Sales  \&
  Torrey}{Li et~al.}{2020}]{liEffectsSubgridModels2020}
Li H.,  Vogelsberger M.,  Marinacci F.,  Sales L.~V.,   Torrey P.,  2020,
  \mn@doi [Monthly Notices of the Royal Astronomical Society]
  {10.1093/mnras/staa3122}, 499, 5862

\bibitem[\protect\citeauthoryear{Lin et~al.,}{Lin
  et~al.}{2019}]{linSDSSIVMaNGAInsideout2019}
Lin L.,  et~al., 2019, \mn@doi [The Astrophysical Journal]
  {10.3847/1538-4357/aafa84}, 872, 50

\bibitem[\protect\citeauthoryear{Lotz, Jonsson, Cox, Croton, Primack,
  Somerville  \& Stewart}{Lotz et~al.}{2011}]{lotzMajorMinorGalaxy2011}
Lotz J.~M.,  Jonsson P.,  Cox T.~J.,  Croton D.,  Primack J.~R.,  Somerville
  R.~S.,   Stewart K.,  2011, \mn@doi [The Astrophysical Journal]
  {10.1088/0004-637X/742/2/103}, 742, 103

\bibitem[\protect\citeauthoryear{Lu, Mo, Katz  \& Weinberg}{Lu
  et~al.}{2006}]{luOriginColdDark2006}
Lu Y.,  Mo H.~J.,  Katz N.,   Weinberg M.~D.,  2006, \mn@doi [Monthly Notices
  of the Royal Astronomical Society] {10.1111/j.1365-2966.2006.10270.x}, 368,
  1931

\bibitem[\protect\citeauthoryear{Ma et~al.,}{Ma
  et~al.}{2020}]{maSelfconsistentProtoglobularCluster2020}
Ma X.,  et~al., 2020, \mn@doi [Monthly Notices of the Royal Astronomical
  Society] {10.1093/mnras/staa527}, 493, 4315

\bibitem[\protect\citeauthoryear{Macci{\`o}, Dutton, Van Den~Bosch, Moore,
  Potter  \& Stadel}{Macci{\`o}
  et~al.}{2007}]{maccioConcentrationSpinShape2007}
Macci{\`o} A.~V.,  Dutton A.~A.,  Van Den~Bosch F.~C.,  Moore B.,  Potter D.,
  Stadel J.,  2007, \mn@doi [Monthly Notices of the Royal Astronomical Society]
  {10.1111/j.1365-2966.2007.11720.x}, 378, 55

\bibitem[\protect\citeauthoryear{McCluskey, Wetzel, Loebman, Moreno  \&
  {Faucher-Giguere}}{McCluskey
  et~al.}{2023}]{mccluskeyDiskSettlingDynamical2023}
McCluskey F.,  Wetzel A.,  Loebman S.~R.,  Moreno J.,   {Faucher-Giguere}
  C.-A.,  2023, Disk Settling and Dynamical Heating: Histories of {{Milky
  Way-mass}} Stellar Disks across Cosmic Time in the {{FIRE}} Simulations,
  \mn@doi{10.48550/arXiv.2303.14210}

\bibitem[\protect\citeauthoryear{Miller, van Dokkum, Mowla  \& van~der
  Wel}{Miller et~al.}{2019}]{millerNewViewSize2019}
Miller T.~B.,  van Dokkum P.,  Mowla L.,   van~der Wel A.,  2019, \mn@doi
  [ApJL] {10.3847/2041-8213/ab0380}, 872, L14

\bibitem[\protect\citeauthoryear{Mishra, Rana  \& More}{Mishra
  et~al.}{2023}]{mishraStellarMassDependence2023}
Mishra P.~K.,  Rana D.,   More S.,  2023, \mn@doi [Monthly Notices of the Royal
  Astronomical Society] {10.1093/mnras/stad2914}, 526, 2403

\bibitem[\protect\citeauthoryear{Mo, Mao  \& White}{Mo
  et~al.}{1998}]{moFormationGalacticDiscs1998}
Mo H.~J.,  Mao S.,   White S. D.~M.,  1998, \mn@doi [Monthly Notices of the
  Royal Astronomical Society] {10.1046/j.1365-8711.1998.01227.x}, 295, 319

\bibitem[\protect\citeauthoryear{Mo, Chen  \& Wang}{Mo
  et~al.}{2023}]{moTwophaseModelGalaxy2023}
Mo H.,  Chen Y.,   Wang H.,  2023, A Two-Phase Model of Galaxy Formation:
  {{I}}. {{The}} Growth of Galaxies and Supermassive Black Holes (\mn@eprint
  {arxiv} {2311.05030}), \mn@doi{10.48550/arXiv.2311.05030}

\bibitem[\protect\citeauthoryear{Mowla, van~der Wel, van Dokkum  \&
  Miller}{Mowla et~al.}{2019}]{mowlaMassdependentSlopeGalaxy2019}
Mowla L.,  van~der Wel A.,  van Dokkum P.,   Miller T.~B.,  2019, \mn@doi
  [ApJL] {10.3847/2041-8213/ab0379}, 872, L13

\bibitem[\protect\citeauthoryear{Murray}{Murray}{2014}]{murrayHMFHaloMass2014}
Murray S.,  2014, Astrophysics Source Code Library, p. ascl:1412.006

\bibitem[\protect\citeauthoryear{Naab, Johansson  \& Ostriker}{Naab
  et~al.}{2009}]{naabMinorMergersSize2009}
Naab T.,  Johansson P.~H.,   Ostriker J.~P.,  2009, \mn@doi [The Astrophysical
  Journal] {10.1088/0004-637X/699/2/L178}, 699, L178

\bibitem[\protect\citeauthoryear{Navarro, Frenk  \& White}{Navarro
  et~al.}{1997}]{navarroUniversalDensityProfile1997}
Navarro J.~F.,  Frenk C.~S.,   White S. D.~M.,  1997, \mn@doi [ApJ]
  {10.1086/304888}, 490, 493

\bibitem[\protect\citeauthoryear{Nedkova et~al.,}{Nedkova
  et~al.}{2021}]{nedkovaExtendingEvolutionStellar2021}
Nedkova K.~V.,  et~al., 2021, \mn@doi [Monthly Notices of the Royal
  Astronomical Society] {10.1093/mnras/stab1744}, 506, 928

\bibitem[\protect\citeauthoryear{Newman, Ellis, Bundy  \& Treu}{Newman
  et~al.}{2012}]{newmanCanMinorMerging2012}
Newman A.~B.,  Ellis R.~S.,  Bundy K.,   Treu T.,  2012, \mn@doi [The
  Astrophysical Journal] {10.1088/0004-637X/746/2/162}, 746, 162

\bibitem[\protect\citeauthoryear{Oser, Ostriker, Naab, Johansson  \&
  Burkert}{Oser et~al.}{2010}]{oserTwoPhasesGalaxy2010}
Oser L.,  Ostriker J.~P.,  Naab T.,  Johansson P.~H.,   Burkert A.,  2010,
  \mn@doi [The Astrophysical Journal] {10.1088/0004-637X/725/2/2312}, 725, 2312

\bibitem[\protect\citeauthoryear{Oser, Naab, Ostriker  \& Johansson}{Oser
  et~al.}{2012}]{oserCosmologicalSizeVelocity2012}
Oser L.,  Naab T.,  Ostriker J.~P.,   Johansson P.~H.,  2012, \mn@doi [The
  Astrophysical Journal] {10.1088/0004-637X/744/1/63}, 744, 63

\bibitem[\protect\citeauthoryear{Perez \& Granger}{Perez \&
  Granger}{2007}]{perezIPythonSystemInteractive2007}
Perez F.,  Granger B.~E.,  2007, \mn@doi [Computing in Science and Engineering]
  {10.1109/MCSE.2007.53}, 9, 21

\bibitem[\protect\citeauthoryear{{Planck Collaboration} et~al.,}{{Planck
  Collaboration} et~al.}{2016}]{planckcollaborationPlanck2015Results2016}
{Planck Collaboration} et~al., 2016, \mn@doi [Astronomy and Astrophysics]
  {10.1051/0004-6361/201525830}, 594, A13

\bibitem[\protect\citeauthoryear{Robitaille et~al.,}{Robitaille
  et~al.}{2013}]{robitailleAstropyCommunityPython2013}
Robitaille T.~P.,  et~al., 2013, \mn@doi [A\&A] {10.1051/0004-6361/201322068},
  558, A33

\bibitem[\protect\citeauthoryear{{Saha, K.}, {Martinez-Valpuesta, I.}  \&
  {Gerhard, O.}}{{Saha, K.} et~al.}{2012}]{sahak.DynamicalEvolutionBulge2012}
{Saha, K.} {Martinez-Valpuesta, I.}  {Gerhard, O.} 2012, \mn@doi [EPJ Web of
  Conferences] {10.1051/epjconf/20121906008}, 19, 06008

\bibitem[\protect\citeauthoryear{Saha, Tseng  \& Taam}{Saha
  et~al.}{2010}]{sahaEffectBarsTransient2010}
Saha K.,  Tseng Y.-H.,   Taam R.~E.,  2010, \mn@doi [The Astrophysical Journal]
  {10.1088/0004-637X/721/2/1878}, 721, 1878

\bibitem[\protect\citeauthoryear{Schaye et~al.,}{Schaye
  et~al.}{2015}]{schayeEAGLEProjectSimulating2015}
Schaye J.,  et~al., 2015, \mn@doi [Monthly Notices of the Royal Astronomical
  Society] {10.1093/mnras/stu2058}, 446, 521

\bibitem[\protect\citeauthoryear{Shen, Mo, White, Blanton, Kauffmann, Voges,
  Brinkmann  \& Csabai}{Shen
  et~al.}{2003}]{shenSizeDistributionGalaxies2003MassSizeRelation}
Shen S.,  Mo H.~J.,  White S. D.~M.,  Blanton M.~R.,  Kauffmann G.,  Voges W.,
  Brinkmann J.,   Csabai I.,  2003, \mn@doi [Monthly Notices of the Royal
  Astronomical Society] {10.1046/j.1365-8711.2003.06740.x}, 343, 978

\bibitem[\protect\citeauthoryear{Shi, Kremer, Grudi{\'c}, {Gerling-Dunsmore}
  \& Hopkins}{Shi et~al.}{2023}]{shiHyperEddingtonBlackHole2023}
Shi Y.,  Kremer K.,  Grudi{\'c} M.~Y.,  {Gerling-Dunsmore} H.~J.,   Hopkins
  P.~F.,  2023, \mn@doi [Monthly Notices of the Royal Astronomical Society]
  {10.1093/mnras/stac3245}, 518, 3606

\bibitem[\protect\citeauthoryear{Somerville et~al.,}{Somerville
  et~al.}{2008}]{somervilleExplanationObservedWeak2008}
Somerville R.~S.,  et~al., 2008, \mn@doi [The Astrophysical Journal]
  {10.1086/523661}, 672, 776

\bibitem[\protect\citeauthoryear{Somerville et~al.,}{Somerville
  et~al.}{2018}]{somervilleRelationshipGalaxyDark2018}
Somerville R.~S.,  et~al., 2018, \mn@doi [Monthly Notices of the Royal
  Astronomical Society] {10.1093/mnras/stx2040}, 473, 2714

\bibitem[\protect\citeauthoryear{Suess, Kriek, Price  \& Barro}{Suess
  et~al.}{2019}]{suessHalfmassRadii70002019}
Suess K.~A.,  Kriek M.,  Price S.~H.,   Barro G.,  2019, \mn@doi [The
  Astrophysical Journal] {10.3847/1538-4357/ab1bda}, 877, 103

\bibitem[\protect\citeauthoryear{Virtanen et~al.,}{Virtanen
  et~al.}{2020}]{virtanenSciPyFundamentalAlgorithms2020}
Virtanen P.,  et~al., 2020, \mn@doi [Nat Methods] {10.1038/s41592-019-0686-2},
  17, 261

\bibitem[\protect\citeauthoryear{Wechsler, Bullock, Primack, Kravtsov  \&
  Dekel}{Wechsler et~al.}{2002}]{wechslerConcentrationsDarkHalos2002}
Wechsler R.~H.,  Bullock J.~S.,  Primack J.~R.,  Kravtsov A.~V.,   Dekel A.,
  2002, \mn@doi [ApJ] {10.1086/338765}, 568, 52

\bibitem[\protect\citeauthoryear{Weinberger et~al.,}{Weinberger
  et~al.}{2017}]{weinbergerSimulatingGalaxyFormation2017}
Weinberger R.,  et~al., 2017, \mn@doi [Monthly Notices of the Royal
  Astronomical Society] {10.1093/mnras/stw2944}, 465, 3291

\bibitem[\protect\citeauthoryear{Weinberger et~al.,}{Weinberger
  et~al.}{2018}]{weinbergerSupermassiveBlackHoles2018}
Weinberger R.,  et~al., 2018, \mn@doi [Monthly Notices of the Royal
  Astronomical Society] {10.1093/mnras/sty1733}, 479, 4056

\bibitem[\protect\citeauthoryear{Yang, Mo  \& van~den Bosch}{Yang
  et~al.}{2003}]{yangConstrainingGalaxyFormation2003}
Yang X.,  Mo H.~J.,   van~den Bosch F.~C.,  2003, \mn@doi [Monthly Notices of
  the Royal Astronomical Society] {10.1046/j.1365-8711.2003.06254.x}, 339, 1057

\bibitem[\protect\citeauthoryear{Yang, Mo, van~den Bosch, Zhang  \& Han}{Yang
  et~al.}{2012}]{yangEVOLUTIONGALAXYDARK2012}
Yang X.,  Mo H.~J.,  van~den Bosch F.~C.,  Zhang Y.,   Han J.,  2012, \mn@doi
  [ApJ] {10.1088/0004-637X/752/1/41}, 752, 41

\bibitem[\protect\citeauthoryear{Zhao, Mo, Jing  \& B{\"o}rner}{Zhao
  et~al.}{2003}]{zhaoGrowthStructureDark2003}
Zhao D.~H.,  Mo H.~J.,  Jing Y.~P.,   B{\"o}rner G.,  2003, \mn@doi [Monthly
  Notices of the Royal Astronomical Society]
  {10.1046/j.1365-8711.2003.06135.x}, 339, 12

\bibitem[\protect\citeauthoryear{Zhuang \& Ho}{Zhuang \&
  Ho}{2023}]{zhuangEvolutionaryPathsActive2023}
Zhuang M.-Y.,  Ho L.~C.,  2023, \mn@doi [Nat Astron]
  {10.1038/s41550-023-02051-4}, pp 1--14

\bibitem[\protect\citeauthoryear{{van Dokkum} et~al.,}{{van Dokkum}
  et~al.}{2008}]{vandokkumConfirmationRemarkableCompactness2008}
{van Dokkum} P.~G.,  et~al., 2008, \mn@doi [The Astrophysical Journal]
  {10.1086/587874}, 677, L5

\bibitem[\protect\citeauthoryear{{van Dokkum} et~al.,}{{van Dokkum}
  et~al.}{2015}]{vandokkumFormingCompactMassive2015}
{van Dokkum} P.~G.,  et~al., 2015, \mn@doi [The Astrophysical Journal]
  {10.1088/0004-637X/813/1/23}, 813, 23

\bibitem[\protect\citeauthoryear{{van der Wel} et~al.,}{{van der Wel}
  et~al.}{2014}]{vanderwel3DHSTCANDELSEvolution2014}
{van der Wel} A.,  et~al., 2014, \mn@doi [The Astrophysical Journal]
  {10.1088/0004-637X/788/1/28}, 788, 28

\bibitem[\protect\citeauthoryear{{van der Wel} et~al.,}{{van der Wel}
  et~al.}{2024}]{vanderwelStellarHalfmassRadii2024}
{van der Wel} A.,  et~al., 2024, \mn@doi [The Astrophysical Journal]
  {10.3847/1538-4357/ad02ee}, 960, 53

\makeatother
\end{thebibliography}



\appendix
 
\section{Star formation effficency} \label{sec:app:sfe}

\begin{figure} \centering
    \includegraphics[width=0.975\columnwidth]{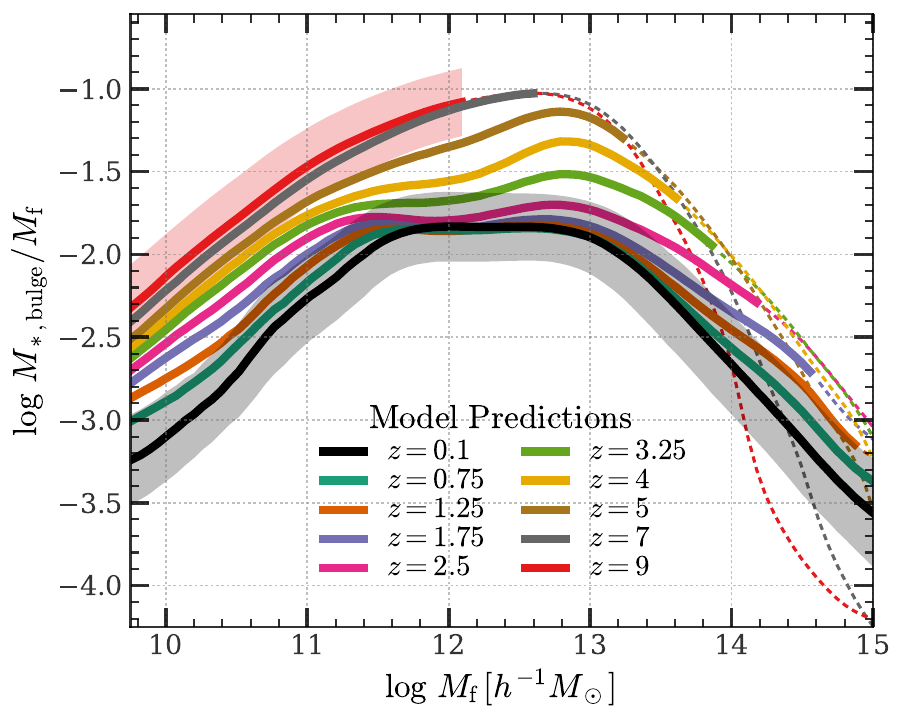}
    \caption{
    The ratio of bulge stellar mass ($M_{\rm *,bulge}$) to halo mass assembled 
    in the fast phase ($M_{\rm f}$) as a function of $M_{\rm f}$, 
    predicted by our model at different redshifts. 
    Here, dynamically hot galaxies, defined as those with a bulge stellar 
    mass fraction $f_{\rm bulge} > 0.75$ (see \S\ref{ssec:compare-with-obs}), 
    are included in the analysis. The {\bf curves} represent median relations, 
    while {\bf shaded areas} indicate the 16th-84th percentile ranges 
    (shown only for the two extreme redshifts for clarity). The {\bf dashed tail} 
    of each curve marks the range of halo mass where the halo mass function is 
    lower than $1.0\, h^{3}{\rm Gpc}^{-3}{\rm dex}^{-1}$ and thus the 
    hosted galaxies are too rare to be observed.
    The plateau feature of these curves is 
    inferred as the reason for the $r_{\rm bulge}$-$M_{\rm *,bulge}$ relation to 
    follow the $r_{\rm f}$-$M_{\rm f}$ relation 
    at $M_{\rm f} \approx 10^{12} \Msun$
    (see \S\ref{ssec:interp-size-mass}). 
    }
    \label{fig:ms2mf_ratio_vs_mf}
\end{figure}

In \S\ref{ssec:interp-size-mass}, we have interpreted the non-linear size-mass 
($r_{\rm bulge}$-$M_{\rm *,bulge}$) relation of dynamically hot galaxies as 
a combined consequence of the near-linear halo size-halo mass 
($r_{\rm f}$-$M_{\rm f}$) relation, the homologous bulge size-halo size 
relation ($r_{\rm bulge} \approx 100\, r_{\rm f}$), and the non-linear 
bulge mass-halo mass ($M_{\rm *,bulge}$-$M_{\rm f}$) relation.
Specifically, the follow-up of the $r_{\rm bulge}$-$M_{\rm *,bulge}$ relation 
to the $r_{\rm f}$-$M_{\rm f}$ relation appears at the halo mass where 
the star formation efficiency ($M_{\rm *,bulge}/M_{\rm f}$) reaches its peak, 
and the non-linear flattening and upturn of the $r_{\rm bulge}$-$M_{\rm *,bulge}$ 
relation are produced by the suppression of star formation efficiency at lower and 
higher masses, respectively. Fig.~\ref{fig:ms2mf_ratio_vs_mf} shows the star formation 
efficiency as a function of $M_{\rm f}$ predicted by our model at various redshifts. 
A plateau is seen at $M_{\rm f} \approx 10^{12} \Msun$ at $z \approx 0.1$, 
and it becomes slightly tilted at $z \approx 3$. This feature is expected 
and thus supports our interpretation of the non-linear size-mass relation of 
dynamically hot galaxies.


\bsp	
\label{lastpage}
\end{document}